\documentclass[journal]{IEEEtran}

\ifx\pdfoutput\undefined
\usepackage{verbatim}
\usepackage{graphicx} \else
\usepackage[pdftex]{graphicx} \fi
\usepackage{subfigure}
\usepackage{amsmath}
\usepackage{dsfont}
\usepackage{amssymb}
\usepackage{amsthm}
\usepackage{algorithm}
\usepackage{subfigure}
\usepackage{cuted}
\usepackage{epstopdf}
\usepackage{cite}
\usepackage{soul,color,bm}
\usepackage{setspace}
\usepackage{amsbsy}
\usepackage[utf8]{inputenc}
\usepackage{booktabs,lipsum}
\usepackage{algpseudocode}
\usepackage{comment}
\usepackage{xcolor}

\newtheorem{thm}{Theorem}

\newtheorem{prop}{Proposition}

\usepackage{lipsum} 
\usepackage{multicol}
\usepackage{multirow}
\usepackage{float}
\usepackage{graphicx}

\usepackage{etoolbox}
\makeatletter
\patchcmd{\@begintheorem}{\textit}{\textbf}{}{}
\makeatother
\ifCLASSINFOpdf
\else
\fi

\begin{document}

\title{Blockage-Aware UAV-Assisted Wireless Data Harvesting With Building Avoidance}

\author{Gitae Park, Kanghyun Heo, and Kisong Lee,~\IEEEmembership{Senior Member,~IEEE}

\thanks{The authors are with the Department of Information and Communication Engineering, Dongguk University, Seoul 04620, South Korea (e-mail: kslee851105@gmail.com).}

\thanks{G. Park and K. Heo are co-first authors and K. Lee is a corresponding author.}
}

\maketitle

\begin{abstract}
Unmanned aerial vehicles (UAVs) offer dynamic trajectory control, enabling them to avoid obstacles and establish line-of-sight (LoS) wireless channels with ground nodes (GNs), unlike traditional ground-fixed base stations. This study addresses the joint optimization of scheduling and three-dimensional (3D) trajectory planning for UAV-assisted wireless data harvesting. The objective is to maximize the minimum uplink throughput among GNs while accounting for signal blockages and building avoidance. To achieve this, we first present mathematical models designed to avoid cuboid-shaped buildings and to determine wireless signal blockage by buildings through rigorous mathematical proof. The optimization problem is formulated as nonconvex mixed-integer nonlinear programming and solved using advanced techniques. Specifically, the problem is decomposed into convex subproblems via quadratic transform and successive convex approximation. Building avoidance and signal blockage constraints are incorporated using the separating hyperplane method and an approximated indicator function. These subproblems are then iteratively solved using the block coordinate descent algorithm. Simulation results validate the effectiveness of the proposed approach. The UAV dynamically adjusts its trajectory and scheduling policy to maintain LoS channels with GNs, significantly enhancing network throughput compared to existing schemes. Moreover, the trajectory of the UAV adheres to building avoidance constraints for its continuous trajectory, ensuring uninterrupted operation and compliance with safety requirements.
\end{abstract}


\IEEEpeerreviewmaketitle

\section{Introduction}

Unmanned aerial vehicles (UAVs) have emerged as a promising technology to enhance next-generation wireless networks due to their high flexibility and cost-effective deployment \cite{Lin18,Wu21}. Unlike traditional ground-fixed base stations (BSs), UAVs can function as mobile BSs, establishing line-of-sight (LoS) air-to-ground (A2G) channels to improve communication capacity \cite{Khuwaja2018}. Consequently, early research has focused on optimizing UAV placement, trajectory, and resource allocation under the assumption of LoS wireless channels between UAVs and ground nodes (GNs) \cite{Fan18,Eom20A,Kang20,Wu18,Heo24,Zhou19,Park23,Kim24,Heo24-3}. 

For instance, the UAV position and communication resources were jointly optimized in \cite{Fan18} to maximize throughput in a UAV relay system. The authors of \cite{Eom20A} addressed the minimum average rate maximization problem considering UAV propulsion energy consumption. Joint trajectory design and resource allocation for simultaneous wireless information and power transfer were investigated in \cite{Kang20,Heo24-3}. In \cite{Park23}, a wireless-powered two-way communication system was proposed, where the UAV broadcasts control signals while GNs receive information and collect energy simultaneously, and each GN uses the harvested energy to transmit data to the UAV. Extensions to multi-UAV scenarios were explored in \cite{Wu18,Kim24}, where scheduling, power control, and UAV trajectories were optimized to account for co-channel interference. Cooperative strategies between BS-UAV and jammer-UAV were also proposed to enhance secure communications against eavesdropping \cite{Heo24,Zhou19}. 

Despite the recognized advantage of LoS channel formation, this assumption has limitations in urban environments dominated by tall and dense buildings. Such environments introduce complexities like multi-path fading and shadowing, making simplistic LoS models insufficient. To address these challenges, a probabilistic LoS channel model has been proposed, which statistically characterizes the likelihood of LoS and non-LoS (NLoS) states based on the elevation angle between the UAV and GNs \cite{Al-Hourani14}. Building on this model, several studies have explored UAV communications by adopting the probabilistic LoS channel model \cite{Zeng19,You20,Duo20,Luo21,Meng22,Duo21,Duo21-1}. For example, \cite{Zeng19} examined UAV energy consumption minimization, including propulsion and communication-related energy. Joint three-dimensional (3D) trajectory and scheduling designs were investigated to maximize the minimum data collection rate for UAV-enabled wireless sensor networks (WSNs) in \cite{You20,Meng22}, with \cite{Meng22} additionally considering energy constraints. Anti-jamming 3D UAV trajectories for legitimate communications were proposed in \cite{Duo20}. Multi-UAV scenarios with joint time allocation and 3D trajectory optimization were examined in \cite{Luo21}, considering the energy-harvesting capabilities of GNs. Furthermore, UAV-enabled jamming strategies to maximize secrecy rates under probabilistic LoS channel models were developed in \cite{Duo21,Duo21-1}.

Although the probabilistic LoS channel model accounts for dense urban environments, it is impractical for real-world applications because it does not consider real-time 3D building characteristics or fixed building locations. Recently, research has shifted toward addressing wireless signal blockage caused by cuboid-shaped buildings \cite{Cai22,Yi22,Yi24,Yi24-2}. For example, geographic information was used in \cite{Yi22} to address A2G link blockages between UAVs and GNs. UAV trajectory and resource allocation optimization to guarantee LoS channels were proposed in \cite{Cai22}. Studies in \cite{Yi24} and \cite{Yi24-2} optimized UAV trajectories and resource allocation to maximize the minimum achievable rate and minimize mission completion time, respectively. However, these studies relied on models that oversimplify building constraints by assuming that UAVs always operate above building heights. As a result, they neither account for building avoidance nor define LoS areas for UAVs operating at lower altitudes. While some research has examined avoiding cylindrical no-fly zones \cite{Li18,Li20,Gao19,Heo24-2}, these approaches do not apply to cuboid-shaped buildings.

While extensive research has explored UAVs under LoS \cite{Fan18,Eom20A,Kang20,Wu18,Heo24,Zhou19,Park23,Kim24,Heo24-3} and probabilistic LoS channel models \cite{Al-Hourani14,Zeng19,You20,Duo20,Luo21,Meng22,Duo21,Duo21-1}, and more recently, signal blockage by buildings \cite{Cai22,Yi22,Yi24,Yi24-2}, no study has developed a generalized channel state determination method applicable across all UAV altitudes that also ensures building avoidance. To address these gaps, this study presents a new joint optimization framework for blockage-aware UAV-assisted wireless data harvesting with building avoidance. The key contributions are summarized as follows: 

\begin{itemize}
     \item We address the joint optimization of scheduling and 3D trajectory planning while accurately determining the LoS/NLoS channel state for UAV-assisted wireless data harvesting in environments with multiple buildings. Specifically, we propose a novel constraint, rigorously proven mathematically, that ensures the UAV avoids cuboid-shaped buildings throughout its continuous trajectory. Additionally, we present a mathematical model to determine whether a wireless signal is blocked by buildings, thereby identifying whether the channel is LoS or NLoS. To the best of our knowledge, this is the first model to simultaneously incorporate building avoidance and generalized channel state determination that accounts for signal blockage, a challenge not tackled in prior research.
     
     \item In the presence of multiple buildings, we formulate an optimization problem to derive the optimal UAV scheduling and 3D trajectory to maximize the minimum throughput among GNs while accounting for building avoidance and wireless signal blockage. This problem, categorized as nonconvex mixed-integer nonlinear programming (MINLP), is solved by decomposing it into subproblems that are convex for specific optimization variables. To achieve this, we employ quadratic transform (QT) and successive convex approximation (SCA) techniques. Furthermore, we introduce novel mathematical methods, including the separating hyperplane method and an approximated indicator function to handle building avoidance and signal blockage constraints effectively. These subproblems are then solved sequentially using an iterative approach based on the block coordinate descent (BCD) algorithm.

     \item Extensive simulations across various scenarios demonstrate that the proposed building avoidance constraint ensures the UAV never encroaches on buildings throughout its continuous trajectory. The UAV effectively forms LoS channels by optimizing its trajectory and scheduling policies, enabling efficient data collection from scheduled GNs. Moreover, the proposed scheme significantly enhances the uplink throughput of GNs compared to baseline approaches.
\end{itemize}

The remainder of this paper is structured as follows. Section II introduces the system model and problem statement. Section III presents the mathematical model for LoS and NLoS state determination. Section IV describes the proposed iterative approach using advanced optimization techniques. Section V evaluates the performance of the proposed scheme and discusses UAV strategies. Finally, Section VI concludes the paper with key insights.

\section{System Model and Problem Formulation}

\begin{figure}[t]
\centering
\includegraphics[width=0.9\linewidth]{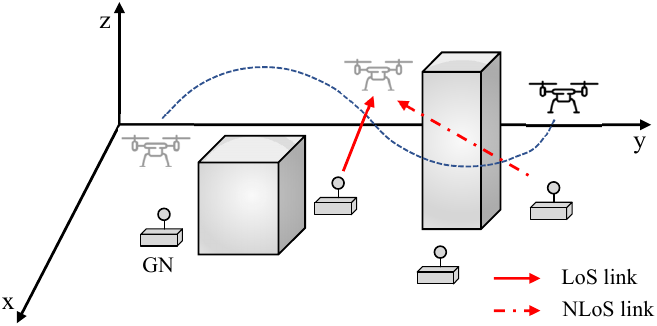} \caption{System model of a UAV-assisted wireless communication network.}
\label{fig1}
\end{figure}

As shown in Fig. \ref{fig1}, we consider a UAV-assisted wireless communication network where a single UAV collects data from $K$ GNs, indexed by $k \in \mathcal{K}=\{1,2,\cdots,K\}$. Let $T$ represent the UAV's flight period divided into $N$ time slots of equal length with $\delta = \frac{T}{N}$, and indexed by $n\in \mathcal{N}=\{1,2,\cdots,N\}$. The position of the UAV is also assumed to be approximately unchanged within each time slot, given sufficiently small $\delta$ \cite{Wu18}. The frequency spectrum is allocated to the GNs for uplink transmission using time-division multiple access.  

The 3D coordinates of the UAV at time slot $n$ are denoted by $\mathbf{q}[n]=(x[n],y[n],z[n])$, while the fixed 3D coordinates of GN $k$ are given by $\mathbf{w}_k=(x_k,y_k,z_k)$. The UAV starts from an initial location $\mathbf{q}_I$, operates at an altitude of within the allowable range, $H_{\textrm{min}} \leq z[n] \leq H_{\textrm{max}}$, to collect data, and returns to a final location $\mathbf{q}_F$. Let $V_{\textrm{max}}$ and $V_{z}$ represent the maximum flight speed of the UAV in 3D space and the vertical direction, respectively, with $V_{\textrm{max}} > V_{z}$ \cite{You20}. Therefore, the maximum distance the UAV can fly in 3D space and vertical direction during each time slot is limited to $\delta V_{\textrm{max}}$ and $\delta V_{z}$, respectively. The mobility constraints on the UAV can thus be summarized as: 
\begin{align}
&\mathbf{q}[0] = \mathbf{q}_I, ~~ \mathbf{q}[N] = \mathbf{q}_F, \label{constM-1}\\ 
&\|\mathbf{q}[n]-\mathbf{q}[n\!-\!1]\| \leq \delta V_{\textrm{max}}, ~~ \forall n,  \label{constM-3} \\
&|z[n]-z[n\!-\!1]| \leq \delta V_{z} , ~~ \forall n,  \label{constM-4}\\
& H_{\textrm{min}} \leq z[n] \leq H_{\textrm{max}}, ~~ \forall n. \label{constM-5}
\end{align}

Let $s_{k}[n]$ be a binary variable indicating whether GN $k$ is scheduled by the UAV at time slot $n$, i.e., $s_{k}[n] = 1$ if GN $k$ is scheduled for uplink transmission at time slot $n$, and $s_{k}[n] = 0$ otherwise. 
Additionally, the UAV serves at most one GN in each time slot, which can be formulated as:
\begin{align}
&s_{k}[n] \in \{0,1\}, ~~\forall k,n, \label{constS-1} \\
&\sum_{k=1}^{K}s_{k}[n] \leq 1, ~~\forall n. \label{constS-2} 
\end{align}

During the flight, the UAV may encounter $L$ non-overlapping cuboid-shaped buildings and must avoid them. Let the center coordinates of building $l$ be $\mathbf{c}_l = (\mathrm{x}_l, \mathrm{y}_l, 0)$\footnote{Because we do not need to consider the case where the UAV avoids buildings in the --$z$ direction, we simplified the problem by setting $\mathrm{z}_l$ to zero.}, and its width, length, and height are denoted by $\mathcal{W}_l$, $\mathcal{L}_l$ and $\mathcal{H}_l$, respectively. The UAV avoids building $l$ if any of the following constraints are satisfied for every time slot $n$.
\begin{subequations} \label{nffz}
\begin{align}
\left(x[n]-\mathrm{x}_l\right)^2 &\geq \left(\frac{\mathcal{W}_l}{2}\right)^2, \\
\left(y[n]-\mathrm{y}_l\right)^2 &\geq \left(\frac{\mathcal{L}_l}{2}\right)^2, \\
z[n] &\geq \mathcal{H}_l, ~~\forall n,l. 
\end{align}
\end{subequations}

From the fact that the wireless channel between the UAV and GN $k$ at time slot $n$ can be either LoS or NLoS depending on signal blockage by buildings, the channel power gain can be expressed as 
\begin{equation}
\displaystyle
h_{k}[n] =
\left\{\begin{array}{rcl}
h^{\textrm{L}}_{k}[n] \!\!\!&= \displaystyle \frac{\beta_0}{d_{k}[n]^{\alpha_{\textrm{L}}}}, & \mbox{for~LoS,}\\
h^{\textrm{N}}_{k}[n] \!\!\!&= \displaystyle \frac{\mu\beta_0}{d_{k}[n]^{\alpha_{\textrm{N}}}}, & \mbox{for~NLoS,}\\
\end{array}\right. \label{channel}
\end{equation}
where $\beta_0$ is the average channel power gain at a reference distance $1$ m in the LoS state, $\mu < 1$ is the signal attenuation factor for NLoS propagation, $\alpha_{\textrm{L}}$ and $\alpha_{\textrm{N}}$ are the average path-loss exponents for the LoS and NLoS states, respectively, with $\alpha_{\textrm{L}} < \alpha_{\textrm{N}}$ \cite{You20}, and $d_{k}[n]$ is the distance between the UAV and GN $k$ at time slot $n$, defined as $d_{k}[n]=\|\mathbf{q}[n]\!-\!\mathbf{w}_k\|$. 

Let $c_{k}^{\textrm{L}}[n]$ denote the binary LoS indicator, indicating whether the channel between the UAV and GN $k$ at time slot $n$ is LoS, i.e., $c_{k}^{\textrm{L}}[n] = 1$ if the channel is LoS and $c_{k}^{\textrm{L}}[n] = 0$ otherwise. Using $c_{k}^{\textrm{L}}[n]$, the channel power gain in \eqref{channel} can be transformed to the following equivalent form. 
\begin{align}
c_{k}^{\textrm{L}}[n] &\in \{0,1\}, ~~ \forall k,n, \label{constraintc}\\
h_{k}[n] &= c_{k}^{\textrm{L}}[n]h_{k}^{\textrm{L}}[n] \!+\! (1\!-\!c_{k}^{\textrm{L}}[n])h_{k}^{\textrm{N}}[n]. \label{hkc}
\end{align}

Then, the achievable uplink spectral efficiency (SE) of GN $k$ at time slot $n$ is given by  
\begin{align}
r_k[n] &= s_{k}[n]\log_2\left(1+\frac{p_kh_{k}[n]}{\sigma^2}\right), ~~\forall k,n, \label{rk}
\end{align}
where $p_k$ is a constant transmit power of GN $k$, and $\sigma^2$ is the noise power. 

Subsequently, the time-averaged SE of GN $k$ is 
\begin{align}
R_k &= \frac{1}{N}\sum_{n=1}^{N}r_k[n], ~~\forall k. \label{rk2}
\end{align}

In this study, our objective is to maximize the minimum uplink SE among the GNs by properly determining the LoS/NLoS states of the wireless channels while avoiding multiple buildings throughout the UAV's continuous trajectory. To achieve this, we aim to optimize the scheduling $\mathbf{S}\triangleq\{s_{k}[n], ~\forall k,n\}$, the 3D trajectory $\mathbf{Q}\triangleq\{\mathbf{q}[n], ~\forall n\}$, and the LoS indicator $\mathbf{C}\triangleq\{c_{k}^{\textrm{L}}[n], ~\forall k,n\}$. Defining $\bar{R} \!=\! \displaystyle \min_{k\in\mathcal{K}} R_{k}$, we can formulate the optimization problem as follows:
\begin{align}
\textbf{(P0):}
~\max_{\substack{\mathbf{S},~\mathbf{Q},~\mathbf{C},~\bar{R}}} &~~~~~\bar{R} \nonumber\\
\textrm{s. t.} ~~~&~~~ R_{k} \geq \bar{R}, ~~\forall k, \label{constrk} \\
&~~~\eqref{constM-1}\!-\!\eqref{nffz},~\eqref{constraintc}. \nonumber
\end{align}
The optimization problem $\textbf{(P0)}$ is an MINLP because $\mathbf{S}$ and $\mathbf{C}$ are binary variables, and constraints \eqref{nffz} and \eqref{constrk} are not convex sets with respect to (w.r.t.) the related optimization variables. Furthermore, \eqref{hkc} must be converted into a tractable form to optimize $\mathbf{C}$. Consequently, analytically deriving a globally optimal solution to this original problem is challenging.

\section{Channel State Determination and Problem Reformulation}

To determine the value of the LoS indicator $c_{k}^{\textrm{L}}[n]$, we must evaluate whether the wireless channel between the UAV and GN $k$ is blocked by one or more buildings. The channel is considered NLoS if there is an intersection between the wireless channel and any building; otherwise, it is LoS.

The components of the line segment between the UAV to GN $k$, denoted as $\mathbf{q}^t_{k}[n]=(x^t_{k}[n],y^t_{k}[n],z^t_{k}[n])$, can be expressed as 
\begin{align}
    \begin{bmatrix}
         x^t_{k}[n]\\
         y^t_{k}[n]\\
         z^t_{k}[n]
    \end{bmatrix}^T =
    \begin{bmatrix}
         x_{k} + (x[n] - x_{k})t \\
         y_{k} + (y[n] - y_{k})t \\
         z_{k} + (z[n] - z_{k})t
    \end{bmatrix}^T,
    \label{line_segment}
\end{align}
where $0 \leq t \leq 1$ is a continuous value representing the internal division of $\mathbf{q}^t_{k}[n]$ used to evaluate the blockage with building $l$. 

If \emph{any component} of $\mathbf{q}^t_{k}[n]$, such as $x^t_{k}[n]$, $y^t_{k}[n]$, and $z^t_{k}[n]$, falls outside the width, length, and height of building $l$ for \emph{all $0 \leq t \leq 1$}, the wireless channel is not blocked by building $l$. Consequently, the channel is considered LoS if \emph{any} of the following constraints are satisfied for all values of $t$.
\begin{subequations} \label{loss}
\begin{align} 
\left(x^t_{k}[n]-\mathrm{x}_l\right)^2 &\geq \left(\frac{\mathcal{W}_l}{2}\right)^{\!2}, \label{loss1}\\
\left(y^t_{k}[n]-\mathrm{y}_l\right)^2 &\geq \left(\frac{\mathcal{L}_l}{2}\right)^{\!2}, \label{loss2}\\
z^t_{k}[n] &\geq \mathcal{H}_l, ~~ \forall t,k,n,l. \label{loss3}
\end{align}
\end{subequations}
While \eqref{loss} determines the LoS of the channel, if the channel is NLoS, there exists a value of $t$ that does not satisfy these constraints, making them infeasible for the NLoS scenario.


To address this challenge, we introduce a big-$\mathrm{M}$ method and binary auxiliary variables $\beta^{(i),t}_{k,l}[n]$ for $i\in\{1,2,3\}$, making the constraints more tractable. These are defined as follows:
\begin{subequations} \label{dsm0}
\begin{align} 
&\rho_{k,l}[n],~\beta^{(i),t}_{k,l}[n] \in \{0,1\}, \label{nu1}\\
&\left(x^t_{k}[n]-\mathrm{x}_l\right)^2 \geq \left(\frac{\mathcal{W}_l}{2}\right)^{\!2}-M(1-\beta^{(1),t}_{k,l}[n]),\label{dsm2}\\
&\left(y^t_{k}[n]-\mathrm{y}_l\right)^2 \geq \left(\frac{\mathcal{L}_l}{2}\right)^{\!2}-M(1-\beta^{(2),t}_{k,l}[n]),\label{dsm3}\\
&z^t_{k}[n] \geq \mathcal{H}_l-M(1-\beta^{(3),t}_{k,l}[n]), \label{dsm4} \\
&\sum_{i=1}^3 \beta^{(i),t}_{k,l}[n] \geq \rho_{k,l}[n],~\forall t,k,n,l,i, \label{dsm5}
\end{align}
\end{subequations}
where $M$ is a sufficiently large constant, ensuring that $M \gg \{\left(\frac{\mathcal{W}_l}{2}\right)^{\!2},\left(\frac{\mathcal{L}_l}{2}\right)^{\!2},\mathcal{H}_l\}$, and $\rho_{k,l}[n]$ is the LoS indicator for the wireless channel between the UAV and GN $k$ w.r.t. building $l$ at time slot $n$. 

If no value of $t$ satisfies the constraints in \eqref{loss}, we set $\beta^{(i),t}_{k,l}[n]=0$, making the problem feasible under the constraints in \eqref{dsm0}. Thus, if the channel is NLoS, the values of $\beta^{(i),t}_{k,l}[n]$ must be $0$ for a given $t$ to satisfy all constraints in \eqref{dsm0}. Subsequently, the LoS indicator $\rho_{k,l}[n]$ is also set to $0$ according to \eqref{dsm5}. In other words, if the channel is NLoS, $\rho_{k,l}[n]$ must be set to $0$, so \eqref{dsm0} can accurately determine the NLoS state of the wireless channel. 

If the channel is LoS, at least one of the constraints \eqref{dsm2}--\eqref{dsm4} is always satisfied for every $t$, even if the corresponding $\beta^{(i),t}_{k,l}[n]$ is between $0$ and $1$. Nonetheless, $\rho_{k,l}[n]$ is more likely to be set to $1$ when the channel is LoS, as this improves the signal reception from the scheduled GN, maximizing $R_{k}$.

We introduce a slack variable $\bar{c}_{k}^{\textrm{L}}[n]$ to determine whether the wireless channel between the UAV and GN $k$ at time slot $n$ is LoS or NLoS, considering all potential building blockages. When the channel is blocked by one or more buildings, it is considered NLoS. Therefore, $\bar{c}_{k}^{\textrm{L}}[n]$ must be $0$ if $\rho_{k,l}[n]$ is 0 for at least one $l$, leading to the following constraint:
\begin{align}
     0 &\leq \bar{c}_{k}^{\textrm{L}}[n] \leq \rho_{k,l}[n], ~~\forall k,n,l.   \label{nlosd1}
\end{align}
If $\rho_{k,l}[n]=1$ for all $l$, $\bar{c}_{k}^{\textrm{L}}[n]$ is likely to be set to $1$, as it is beneficial to improve $R_{k}$. Considering all constraints in \eqref{dsm0} and \eqref{nlosd1}, we can determine whether the wireless channel is LoS or NLoS for multiple building blockages by examining $\bar{c}_{k}^{\textrm{L}}[n]$. Specifically, $\bar{c}_{k}^{\textrm{L}}[n]=0$ if the channel is NLoS and $\bar{c}_{k}^{\textrm{L}}[n] = 1$ otherwise.

As explained, the big-$\mathrm{M}$ method is primarily useful for determining the NLoS state, as even if a LoS channel is incorrectly identified as NLoS, the lower bound of the signal channel remains guaranteed. Therefore, we can use $\bar{c}_{k}^{\textrm{L}}[n]$ to establish the lower bound of the signal channel, as follows:
\begin{align}
 h_{k}[n] \!\geq\! \bar{c}_{k}^{\textrm{L}}[n]h_{k}^{\textrm{L}}[n] \!+\! (\!1\!-\!\bar{c}_{k}^{\textrm{L}}[n])h_{k}^{\textrm{N}}[n] \!\triangleq\! h_{k}^{\textrm{LB}}[n]. \label{lb1}
 \end{align}

From \eqref{lb1}, the lower bound of the uplink spectral efficiency $r_{k}[n]$ is given by
\begin{align}
r^{\textrm{LB}}_k[n] &= s_{k}[n]\log_2\bigg(1+ \frac{p_kh^{\textrm{LB}}_{k}[n]}{\sigma^2}\bigg). \label{rlbk}
\end{align}
The corresponding time-averaged SE is 
\begin{align}
R^{\textrm{LB}}_k &= \frac{1}{N}\sum_{n=1}^{N}r^{\textrm{LB}}_k[n], ~~\forall k.
\end{align}

Finally, we can reformulate the original problem \textbf{(P0)} into the following tractable form: 
\begin{align}
\textbf{(P1):}
~\max_{\substack{\mathbf{S},~\mathbf{Q},~\bar{\mathbf{C}},~\pmb{\rho},~\pmb{\beta},~\eta}} &~~~~~\eta \nonumber\\
\textrm{s. t.} ~~~~~~& R^{\textrm{LB}}_{k} \geq \eta, ~~\forall k, \label{constrk2} \\
&\eqref{constM-1}\!-\!\eqref{nffz},~\eqref{dsm0},~\eqref{nlosd1}, \nonumber 
\end{align}
where $\bar{\mathbf{C}}\triangleq\{\bar{c}_{k}^{\textrm{L}}[n], ~\forall k,n\}$, $\pmb{\rho}\triangleq\{\rho_{k,l}[n], ~\forall k,n,l\}$, and $\pmb{\beta}\triangleq\{\beta^{(i),t}_{k,l}[n], ~\forall t,k,n,l,i\}$.

\section{Proposed Algorithm}

Problem $\textbf{(P1)}$ remains challenging to solve due to the nonconvex nature of the constraints. To address this, we decompose it into two subproblems and employ SCA and QT to convert each subproblem into a convex form w.r.t. the relevant optimization variables. This enables us to solve the problem using existing convex optimization solvers, such as CVX \cite{Grant}. Additionally, we propose new optimization methods to handle building avoidance and signal blockage constraints. Finally, we develop an iterative algorithm that applies BCD to sequentially solve the relaxed convex problems. The specific procedures for addressing each subproblem are described below.

\subsection{Scheduling Optimization}

Relaxing $s_{k}[n]$ to take continuous values between $0$ and $1$, the problem of finding the optimal $\mathbf{S}$ for fixed values of the remaining variables can be formulated as follows:
\begin{align}
\textbf{(P2):}
~\max_{\mathbf{S},~\eta} &~~~~~~~~\eta \nonumber\\
\textrm{s. t.} &~~~ 0 \leq s_{k}[n] \leq 1, ~~\forall k,n, \\ 
&~~~ \eqref{constS-2},~\eqref{constrk2}. \nonumber
\end{align}
Problem \textbf{(P2)} is a standard linear programming, and can be efficiently solved using CVX. The resulting continuous scheduling solution can subsequently be reconstructed into binary scheduling using the method in \cite{Wu18} without compromising optimality.

\subsection{3D Trajectory and LoS indicator Optimization}

Since the variables $\mathbf{Q}$ and $\bar{\mathbf{C}}$ are closely interrelated, we optimize them simultaneously. For fixed $\mathbf{S}$, the optimization problem is reformulated as follows:
\begin{align}
\textbf{(P3):}
~\max_{\substack{\mathbf{Q},~\bar{\mathbf{C}},~\pmb{\rho},~\pmb{\beta},~\eta}} ~&~~~~~\eta \nonumber\\
\textrm{s. t.} ~~~~~& \eqref{constM-1}\!-\!\eqref{constM-5},~\eqref{nffz},~\eqref{dsm0},~\eqref{nlosd1},~\eqref{constrk2}. \nonumber
\end{align}
Here, constraints \eqref{nffz}, \eqref{dsm0}, and \eqref{constrk2} are required to be handled to make problem \textbf{(P3)} convex.

\begin{figure}[t]
  \begin{center}
    \subfigure[Conventional.]{
      \includegraphics[width=0.4\linewidth]{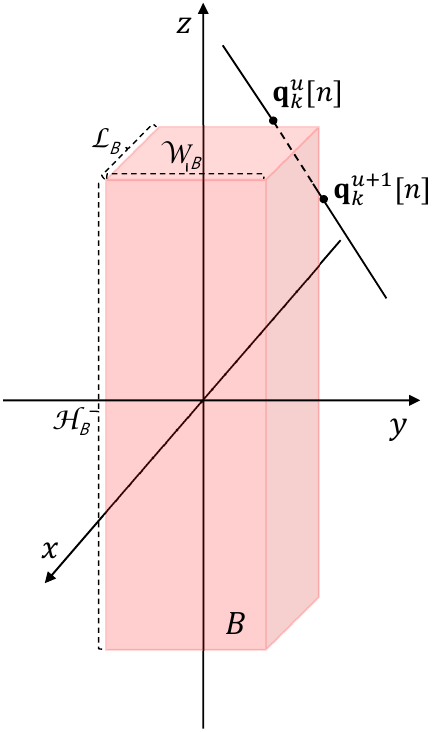}\label{fig2a}
    }
    \subfigure[Proposed.]{
      \includegraphics[width=0.53\linewidth]{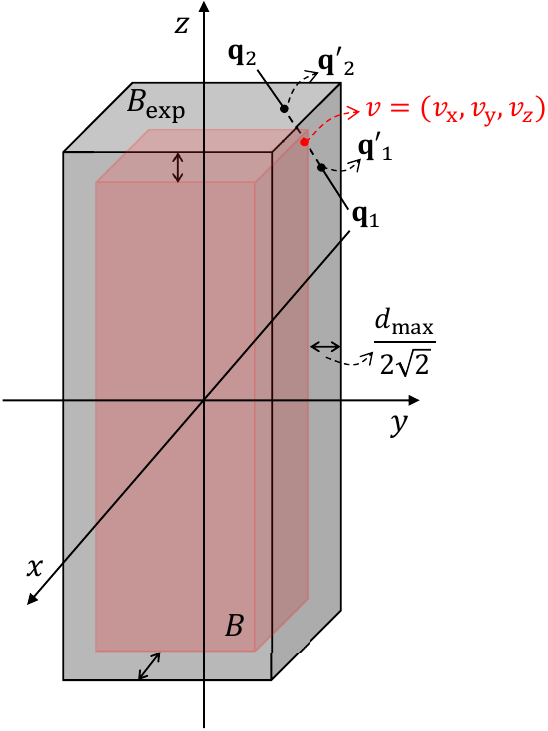}\label{fig2b}
    }
  \end{center} 
\caption{Channel state determination constraints.}
\label{fig2}
\end{figure}

\vspace{2mm}
\subsubsection{Constraint on Channel State Determination \eqref{dsm0}}

Determining whether a channel is LoS or NLoS requires evaluating constraint \eqref{dsm0} for all continuous values of $t$ between $0$ and $1$, which is computationally infeasible. To make this constraint simple and tractable, we divide the line segment between the UAV and GN $k$ into $U$ equal segments. Then, the components of the $u$-th point of $\mathbf{q}^{u}_{k}[n]=(x^{u}_{k}[n],y^{u}_{k}[n],z^{u}_{k}[n])$ can be written as
\begin{align}
   \!\! \begin{bmatrix}
         x^{u}_{k}[n]\\
         y^{u}_{k}[n]\\
         z^{u}_{k}[n]
    \end{bmatrix}^{\!T} \!\!\!=\!\!
    \begin{bmatrix}
         x_{k} \!+\! \displaystyle \frac{(x[n] \!-\! x_{k})u}{U} \\
         y_{k} \!+\! \displaystyle \frac{(y[n] \!-\! y_{k})u}{U} \\
         z_{k} \!+\! \displaystyle \frac{(z[n] \!-\! z_{k})u}{U}
    \end{bmatrix}^{\!T}\!\!\!, ~~ u \!\in\! \{0,1,\cdots,U\}.
    \label{line_segment2}
\end{align}

Let $\beta^{(i),u}_{k,l}[n]$ be the auxiliary variable used for the big-M method for the $u$-th point of $\mathbf{q}^{u}_{k}[n]$. We also relax $\rho_{k,l}[n]$ and $\beta^{(i),u}_{k,l}[n]$ to continuous variables between 0 and 1 and use an indicator function to handle their original binary nature. Then, constraint \eqref{dsm0} can be replaced with the following constraint, such that the channel state can be examined for discrete values of $u$.
\begin{subequations} \label{dsm1}
\begin{align} 
&0 \leq \rho_{k,l}[n],~\beta^{(i),u}_{k,l}[n] \leq 1, \label{nu11}\\
&\left(x^{u}_{k}[n]-\mathrm{x}_l\right)^2 \geq \left(\frac{\mathcal{W}_l}{2}\right)^{\!2}-M(1-\beta^{(1),u}_{k,l}[n]),\label{dsm21}\\
&\left(y^{u}_{k}[n]-\mathrm{y}_l\right)^2 \geq \left(\frac{\mathcal{L}_l}{2}\right)^{\!2}-M(1-\beta^{(2),u}_{k,l}[n]),\label{dsm31}\\
&z^{u}_{k}[n] \geq \mathcal{H}_l-M(1-\beta^{(3),u}_{k,l}[n]), \label{dsm41} \\
&\sum_{i=1}^3 \Phi(\beta^{(i),u}_{k,l}[n]) \geq \rho_{k,l}[n], ~~\forall u,k,n,l,i, \label{dsm61}
\end{align}
\end{subequations}
where $\Phi(x)$ denotes an indicator function, defined as follows:
\begin{align}
    \Phi (x) = 
    \begin{cases}
    1,\hspace{0.5cm} \text{if } x\geq 1,\\
    0,\hspace{0.5cm} \text{if } 0 \leq x < 1.
    \end{cases} 
\end{align}

In \eqref{dsm1}, the wireless channel is judged to be NLoS only if $\beta_{k,l}^{(i),u}[n]$, $\forall i$, are all less than 1 by setting $\rho_{k,l}[n]$ to 0 because $\sum_{i=1}^3\Phi(\beta^{(i),u}_{k,l}[n])$ becomes 0 due to the binary nature of the indicator function, otherwise it is judged to be LoS. Therefore, the LoS/NLoS state of the wireless channel can be determined for the discrete values of $u$. However, as shown in Fig. \ref{fig2a}, this approach cannot guarantee LoS/NLoS determination for continuous line segments connecting between adjacent points. To resolve this issue, we employ \textit{Theorem 1}, which introduces a constraint ensuring that the LoS condition is met across all continuous segments.

\begin{thm} 
Consider a cuboid $B_{\textrm{exp}}$ with half-width $\frac{\mathcal{W}_B}{2}+\frac{d_{\textrm{max}}}{2\sqrt{2}}$, half-length $\frac{\mathcal{L}_B}{2}+\frac{d_{\textrm{max}}}{2\sqrt{2}}$, and half-height $\frac{\mathcal{H}_B}{2}+\frac{d_{\textrm{max}}}{2\sqrt{2}}$. Suppose that any line segment connecting two points that are not interior points of $B_{\textrm{exp}}$ has a length less than or equal to $d_{\textrm{max}} < \min(\mathcal{W}_B,\mathcal{L}_B,\mathcal{H}_B)$. Under this condition, this line segment never intersects the interior of a cuboid $B$ with half-width $\frac{\mathcal{W}_B}{2}$, half-length $\frac{\mathcal{L}_B}{2}$, and half-height $\frac{\mathcal{H}_B}{2}$.
\end{thm}

\textit{Proof}: Please refer to the Appendix. \qed
\vspace{1mm}

According to \textit{Theorem 1}, constraints \eqref{dsm21}--\eqref{dsm41} can be modified by setting $d_{\textrm{max}}$ to $\frac{\|\mathbf{q}[n]\!-\!\mathbf{w}_k\|}{2\sqrt{2}U}$, as follows:
\begin{subequations} \label{dsme}
\begin{align}
&\!\!\!\!(x^{u}_{k}[n]\!-\!\mathrm{x}_l)^{\!2} \!\geq\! \left(\!\frac{\mathcal{W}_l}{2}\!+\!\frac{\|\mathbf{q}[n]\!-\!\mathbf{w}_k\|}{2\sqrt{2}U}\!\right)^{\!\!2}\!\!-\!M\!\!\left(\!1\!-\!\beta^{(1),u}_{k,l}[n]\!\right)\!,\label{dsme1}\\
&\!\!\!\!(y^{u}_{k}[n]\!-\!\mathrm{y}_l)^{\!2} \!\geq\! \left(\!\frac{\mathcal{L}_l}{2}\!+\!\frac{\|\mathbf{q}[n]\!-\!\mathbf{w}_k\|}{2\sqrt{2}U}\!\right)^{\!\!2}\!\!-\!M\!\!\left(\!1\!-\!\beta^{(2),u}_{k,l}[n]\!\right)\!,\label{dsme2}\\
&\!\!\!\!z^{u}_{k}[n] \!\geq\! \left(\!\mathcal{H}_l\!+\!\frac{\|\mathbf{q}[n]\!-\!\mathbf{w}_k\|}{2\sqrt{2}U}\!\right)\!\!-\!M\!\!\left(\!1\!-\!\beta^{(3),u}_{k,l}[n]\!\right), ~~\forall u,k,n,l. \label{dsme3}
\end{align}
\end{subequations}
These constraints ensure that the LoS state of the wireless channel for all continuous values connecting any discrete value of $u$, as illustrated in Fig. \ref{fig2b} where the red and grey cuboids represent the original and expanded buildings, respectively. Therefore, constraints \eqref{dsm21}--\eqref{dsm41} can be replaced with constraint \eqref{dsme}.

Additionally, the indicator function in constraint \eqref{dsm61} needs to be transformed into a tractable form because it is piecewise and discontinuous. Constraints \eqref{dsme1} and \eqref{dsme2} are not convex sets and must be converted into convex sets. 
To address these challenges, the indicator function is approximated using two linear functions: 
\begin{align}
    \Phi_{a}(x) &= 
    \begin{cases}
    \Phi_{a+}(x)=ax-a+1, &\text{if } x\geq \frac{a}{a+1},\\
    \Phi_{a-}(x)=\frac{1}{a}x, &\text{if } 0 \leq x \leq \frac{a}{a+1},
    \end{cases} \nonumber\\
    &=\max\bigg(ax-a+1,~\frac{1}{a}x\bigg),
\end{align}
where $\Phi_{a}(x)$ is equivalent to $\Phi(x)$ as $a \rightarrow \infty$. This function is convex since the maximum of two linear functions is convex. Replacing $\Phi(x)$ with $\Phi_{a}(x)$ in \eqref{dsm61} resolves the discontinuity of the indicator function. Nevertheless, constraint \eqref{dsm61} is still not a convex set, so we need to derive the lower bound of $\Phi_{a}(x)$ using the first-order Taylor expansion at a given point $x^r$, as follows:  
\begin{align}
    \Phi_{a}^{\textrm{LB}}(x) \!&=\! 
    \begin{cases}
    \Phi_{a+}(x^r)\!+\!\Phi'_{a+}(x^r)(x\!-\!x^r), &\text{if } x^r\geq \frac{a}{a+1},\\
    \Phi_{a-}(x^r)\!+\!\Phi'_{a-}(x^r)(x\!-\!x^r), &\text{if } 0 \leq x^r \leq \frac{a}{a+1}.
    \end{cases}
\end{align}

Because $\Phi_{a}^{\textrm{LB}}(x)$ is a continuous linear function w.r.t. $x$, we can replace $\Phi(\beta^{(i),u}_{k,l}[n])$ with $\Phi_{a}^{\textrm{LB}}(\beta^{(i),u}_{k,l}[n])$ in \eqref{dsm61} to make this constraint a tractable convex set, as follows: 
\begin{align} 
\sum_{i=1}^3 \Phi_{a}^{\textrm{LB}}(\beta^{(i),u}_{k,l}[n]) \geq \rho_{k,l}[n], ~~\forall u,k,n,l. \label{indc2}
\end{align}
Using a large value of $a$ initially will cause $\Phi_{a}^{\textrm{LB}}(\beta^{(i),u}_{k,l}[n])$ to resemble an indicator function, significantly limiting the feasible region of $\rho_{k,l}[n]$ and making it difficult to find the optimal value of $\rho_{k,l}[n]$. To address this problem, a small value of $a$ is used initially to allow a wider search space for $\rho_{k,l}[n]$, which is gradually increased during iterations to find the optimal value of $\rho_{k,l}[n]$ satisfying \eqref{indc2}.

Given that constraints \eqref{dsme1} and \eqref{dsme2} are not convex sets but their respective left-hand sides (LHSs) have convex form, we can make them convex sets by applying the first-order Taylor expansions to derive lower bounds of their LHSs, as follows:
\begin{subequations} \label{dsme_t}
\begin{align}
&\!\!\!\!2(x_{k}^{u,r}[n]\!-\!\mathrm{x}_l)(x_{k}^{u}[n]\!-\!x_{k}^{u,r}[n])\!+\!(x_{k}^{u,r}[n]\!-\!\mathrm{x}_l)^2 \nonumber\\ 
&\!\!\!\!\geq \!\left(\!\frac{\mathcal{W}_l}{2}\!+\!\frac{\|\mathbf{q}[n]\!-\!\mathbf{w}_k\|}{2\sqrt{2}U}\!\right)^{\!\!2}\!\!-\!M\!\!\left(\!1\!-\!\beta^{(1),u}_{k,l}[n]\!\right)\!, \label{dsme_t1}\\
&\!\!\!\!2(y_{k}^{u,r}[n]\!-\!\mathrm{y}_l)(y_{k}^{u}[n]\!-\!y_{k}^{u,r}[n])\!+\!(y_{k}^{u,r}[n]\!-\!\mathrm{y}_l)^2  \nonumber \\
&\!\!\!\!\geq \!\left(\!\frac{\mathcal{L}_l}{2}\!+\!\frac{\|\mathbf{q}[n]\!-\!\mathbf{w}_k\|}{2\sqrt{2}U}\!\right)^{\!\!2}\!\!-\!M\!\!\left(\!1\!-\!\beta^{(2),u}_{k,l}[n]\!\right)\!, ~\forall u,k,n,l, \label{dsme_t2}
\end{align}
\end{subequations}
where $x_{k}^{u,r}[n]$ and $y_{k}^{u,r}[n]$ are the values of $x_{k}^{u}[n]$ and $y_{k}^{u}[n]$ updated for the $r$-th iteration, respectively. 

Finally, we can replace constraint \eqref{dsm0} with constraints \eqref{nu11}, \eqref{dsm61}, \eqref{dsme3}, \eqref{indc2}, and \eqref{dsme_t}.

\vspace{2mm}
\subsubsection{Constraint on Building Avoidance \eqref{nffz}}

Similar to constraints \eqref{dsm21}--\eqref{dsm41}, constraint \eqref{nffz} ensures that the UAV does not violate buildings at discrete points, i.e., $\mathbf{q}[n]$ for all $n$. However, it does not guarantee that the UAV avoids buildings along its continuous trajectory. To address this problem, constraint \eqref{nffz} is modified using \textit{Theorem 1} by setting $d_{\textrm{max}}$ to $\delta V_{\textrm{max}}$ as follows:
\begin{subequations} \label{nfzexp}
\begin{align} 
\left(x[n]-\mathrm{x}_l\right)^2 &\geq \left(\frac{\mathcal{W}_l}{2}+\frac{\delta V_{\textrm{max}}}{2\sqrt{2}}\!\right)^{\!2}, \label{nfzexp1}\\
\left(y[n]-\mathrm{y}_l\right)^2 &\geq \left(\frac{\mathcal{L}_l}{2}+\frac{\delta V_{\textrm{max}}}{2\sqrt{2}}\!\right)^{\!2}, \label{nfzexp2}\\
z[n] &\geq \left(\mathcal{H}_l+\frac{\delta V_{\textrm{max}}}{2\sqrt{2}}\!\right), ~~\forall n,l. \label{nfzexp3}
\end{align}
\end{subequations}
Notably, the UAV does not violate buildings even when transitioning from $\mathbf{q}[n]$ to $\mathbf{q}[n\!+\!1]$ under constraint \eqref{nfzexp}. However, as with constraint \eqref{dsme}, the big-M method should be applied to handle this constraint, which requires additional auxiliary variables. Therefore, to ensure that the UAV efficiently avoids buildings with low complexity, we use a separating hyperplane method, as explained in \emph{Proposition 1}.

\begin{prop}
    Let $\mathbf{A}$ and $\mathbf{B}$ be disjoint convex sets, i.e., $\mathbf{A} \cap \mathbf{B} = \emptyset$, and $(a^*,b^*)$ is the optimum of $\min_{a,b} \|a-b\|$ for $a \in \mathbf{A}, b\in \mathbf{B}$. Then, the hyperplane $(a^*-b^*)^T(x-b^*)=0$ separates two convex sets $\mathbf{A}$ and $\mathbf{B}$.
\end{prop}

\textit{Proof}: Please refer to Section 2.5 in \cite{Boyd04}. \qed
\vspace{1mm}

Given that the proposed algorithm iteratively finds the UAV strategy, let $\mathbf{q}^r[n]$ represent the trajectory of the UAV at time slot $n$ during the $r$-th iteration. Define $\Omega_{l}$ as the set of points included in the $l$-th expanded building with half-width $\frac{\mathcal{W}_l}{2}+\frac{\delta V_{\textrm{max}}}{2\sqrt{2}}$, half-length $\frac{\mathcal{L}_l}{2}+\frac{\delta V_{\textrm{max}}}{2\sqrt{2}}$, and height $\mathcal{H}_l+\frac{\delta V_{\textrm{max}}}{2\sqrt{2}}$. The point within the $l$-th expanded building closest to $\mathbf{q}^r[n]$, denoted as $\chi_{l}^r[n]$, can be determined by
\begin{align}
    \chi_{l}^r[n] = \min_{\chi_{l} \in \Omega_{l}} \|\mathbf{q}^r[n]-\chi_{l}\|.
\end{align}

Since $\mathbf{q}^r[n]$ is a feasible solution in the $r$-th iteration, it is not contained in $\Omega_{l}$. 
By \emph{Proposition 1}, a hyperplane tangent to $\chi_{l}^r[n]$ can be derived to separate the space containing the building from the space outside it. This hyperplane for the current trajectory of the UAV $\mathbf{q}[n]$ is expressed as $({\mathbf{q}}^r[n] - \chi_{l}^r[n])^T(\mathbf{q}[n]-\chi_{l}^r[n])=0$. Constraint \eqref{nfzexp} can then be replaced with the following condition:
\begin{align}
    ({\mathbf{q}}^r[n] - \chi_{l}^r[n])^T(\mathbf{q}[n]-\chi_{l}^r[n]) > 0, ~~\forall n,l. \label{NFZ_SH}
\end{align}
If constraint \eqref{NFZ_SH} is satisfied, the UAV avoids entering the space of the $l$-th expanded building. By updating the current trajectory based on the previous value, the UAV ensures compliance with building avoidance requirements. Consequently, constraint \eqref{nfzexp} can be effectively replaced by \eqref{NFZ_SH}.

\vspace{2mm}
\subsubsection{Constraint on Minimum Spectral Efficiency \eqref{constrk2}}

To address the nonconvexity of \eqref{constrk2} w.r.t $\mathbf{q}[n]$ and $\bar{c}_{k}^{\textrm{L}}[n]$, we rewrite $h^{\textrm{LB}}_k[n]$ from \eqref{lb1} as the following equivalent expression.
\begin{align}
h_{k}^{\textrm{LB}}[n] = \beta_0\left(\frac{\bar{c}_k^{\textrm{L}}[n]}{\|\mathbf{q}[n]\!-\!\mathbf{w}_k\|^{\alpha_{\textrm{L}}}}+\frac{\mu(1-\bar{c}_k^{\textrm{L}}[n])}{\|\mathbf{q}[n]\!-\!\mathbf{w}_k\|^{\alpha_{\textrm{N}}}}\right). \label{hklb2}
\end{align}

In this form, the fractional terms, $\frac{\bar{c}_k^{\textrm{L}}[n]}{\|\mathbf{q}[n]\!-\!\mathbf{w}_k\|^{\alpha_{\textrm{L}}}}$ and $\frac{\mu(1-\bar{c}_k^{\textrm{L}}[n])}{\|\mathbf{q}[n]\!-\!\mathbf{w}_k\|^{\alpha_{\textrm{N}}}}$, have a \emph{concave-convex} fractional form. Consequently, we can derive an equivalent subtractive form by applying the QT \cite{Shen18} to $h_{k}^{\textrm{LB}}[n]$, as follows:
\begin{align}
&\bar{h}_k^{\textrm{LB}}[n] = \beta_0\bigg(2\lambda_k[n]\sqrt{\bar{c}_k^{\textrm{L}}[n]}-\lambda_k^2[n]\|\mathbf{q}[n]\!-\!\mathbf{w}_k\|^{\alpha_{\textrm{L}}} \nonumber\\
&~~~~~~+2\kappa_k[n]\sqrt{\mu(1\!-\!\bar{c}_k^{\textrm{L}}[n])}-\kappa_k^2[n]\|\mathbf{q}[n]\!-\!\mathbf{w}_k\|^{\alpha_{\textrm{N}}}\bigg), \label{QT}
\end{align}
where $\lambda_k[n]$ and $\kappa_k[n]$ are auxiliary variables for the QT. 

Notably, $\bar{h}_k^{\textrm{LB}}[n]$ is concave w.r.t. $\mathbf{q}[n]$ and $\bar{c}_{k}^{\textrm{L}}[n]$. Using this formulation, the concave lower bound of $\bar{r}_k^{\textrm{LB}}[n]$ is given as 
\begin{align}
\bar{r}_k^{\textrm{LB}}[n] &= s_k[n]\log_2\left(1+\frac{p_k\bar{h}_k^{\textrm{LB}}[n]}{\sigma^2}\right). \label{rklbb}
\end{align}

Additionally, $\bar{r}_k^{\textrm{LB}}[n]$ is concave w.r.t. auxiliary variables, $\lambda_k[n]$ and $\kappa_k[n]$, for fixed $\mathbf{q}[n]$ and $\bar{c}_{k}^{\textrm{L}}[n]$. The optimal values of these variables can be derived by differentiating $\bar{r}_k^{\textrm{LB}}[n]$ over each auxiliary variable, e.g., $\frac{\partial \bar{r}_k^{\textrm{LB}}[n]}{\partial \lambda_k[n]}=0$ and $\frac{\partial \bar{r}_k^{\textrm{LB}}[n]}{\partial \kappa_k[n]}=0$, as follows:
\begin{align}
\lambda^*_k[n] &= \frac{\sqrt{\bar{c}_k^{\textrm{L}}[n]}}{\|\mathbf{q}[n]\!-\!\mathbf{w}_k\|^{\alpha_{\textrm{L}}}}, \label{lambda}\\
\kappa^*_k[n] &= \frac{\sqrt{\mu(1-\bar{c}_k^{\textrm{L}}[n])}}{\|\mathbf{q}[n]\!-\!\mathbf{w}_k\|^{\alpha_{\textrm{N}}}},  ~~\forall k,n. \label{kappa}
\end{align}

Finally, using \eqref{rklbb}, constraint \eqref{constrk2} can be modified to the following convex set.
\begin{align}
\bar{R}_k^{\textrm{LB}} = \frac{1}{N}\sum_{n=1}^N\bar{r}_k^{\textrm{LB}}[n] \geq \eta, ~~\forall k. \label{Rlb}
\end{align}

\vspace{2mm}
\subsubsection{Problem Transformation}

With the modified convex constraints, problem $\textbf{(P3)}$ can be reformulated as the following convex optimization problem:
\begin{align}
\textbf{(P3-1):}
~\max_{\substack{\mathbf{Q},~\bar{\mathbf{C}},~\pmb{\rho},\\ \bar{\pmb{\beta}}, ~\pmb{\lambda},~\pmb{\kappa},~\eta}} ~&~~~~~\eta \nonumber\\
\textrm{s. t.} ~~~& \eqref{constM-1}\!-\!\eqref{constM-5},~\eqref{nlosd1},~\eqref{NFZ_SH},~\eqref{nu11},~\eqref{dsm61}, \nonumber \\
& \eqref{dsme3},~\eqref{indc2},~\eqref{dsme_t},~\eqref{Rlb}, \nonumber
\end{align}
where $\pmb{\lambda}\!\triangleq\!\{\lambda_{k}[n], ~\forall k,n\}$, $\pmb{\kappa}\!\triangleq\!\{\kappa_{k}[n], ~\forall k,n\}$, and $\bar{\pmb{\beta}}\!\triangleq\!\{\beta^{(i),u}_{k,l}[n], ~\forall u,k,n,l,i\}$.

\vspace{2mm}
\subsection{Procedure of Proposed Algorithm}

Both subproblems, $\textbf{(P2)}$ and $\textbf{(P3-1)}$, are convex w.r.t. their respective optimization variables. These subproblems can be solved using a convex solver until convergence. The detailed procedure is outlined in Algorithm \ref{Alg1}. It ensures convergence as the objective function for each subproblem is non-decreasing and bounded above \cite{Wu18}. By analyzing the computational complexity of the worst-case scenario for the interior point method \cite{Ben-Tal01,Boyd04}, the complexity of the proposed algorithm is derived as $\emph{O}\!\left(R_C(UKNL)^{3.5}\log(1/\epsilon)\right)$, where $R_C$ represents the number of iterations needed for convergence (lines 3--10).

\begin{algorithm}[h]
    \caption{Proposed Algorithm} \label{Alg1} \small
    1:$~$Set $r\!=\!0$ and initialize $\mathbf{S}^{r}$, $\mathbf{Q}^{r}$, $\bar{\mathbf{C}}^r$, $a^r>0$, and $\varepsilon > 1$ \\
    2:$~$Calculate $f^r = \min_{k\in\mathcal{K}}R_{k}$\\
    3:$~$\textbf{repeat}  \\
    4:$~~~$Update $r \leftarrow r+1$  \\
    5:$~~~$Update $f^{\textrm{old}} \leftarrow f^{r-1}$\\
    6:$~~~$Find $\mathbf{S}^{r}$ by solving $\textbf{(P2)}$ for given $\{\mathbf{S}^{r-1}\!,\mathbf{Q}^{r-1}\!,\bar{\mathbf{C}}^{r-1}\}$  \\
    7:$~~~$Update $\{\pmb{\lambda}^{r}, \pmb{\kappa}^{r}\}$ using \eqref{lambda} and \eqref{kappa} \\
    8:$~~~$Find $\{\mathbf{Q}^{r},\bar{\mathbf{C}}^r\}$ by solving $\textbf{(P3-1)}$ for given $\{\mathbf{S}^{r}\!,\mathbf{Q}^{r-1}\!,\bar{\mathbf{C}}^{r-1}\}$\\
    9:$~~~$Update $a^r \leftarrow \varepsilon a^{r-1}$\\
    10:$~~$Calculate $f^{r} = \min_{k\in\mathcal{K}}\bar{R}_k^{\textrm{LB}}$\\
    11:$~$\textbf{until} $|f^{r}-f^{\textrm{old}}| < \epsilon$ 
\end{algorithm}

\section{Simulation Results and Discussions}

\begin{table}[ht]
\begin{center}
\caption{Parameter Setup} \small
\begin{tabular}{ll} \hline 
Description & Value \\ \hline \hline
Number of GNs & $K$ = 4 \\
Number of buildings & $L$ =  2\\
Flight period & $T$ = 25 s\\
Number of time slots & $N$ = 50 \\
Length of time slots & $\delta$ = 0.5 s \\
Minimum altitude & $H_{\textrm{min}}$ = 30 m \\
Maximum altitude & $H_{\textrm{max}}$ = 200 m \\
Maximum flight speed in 3D space & $V_{\textrm{max}}$ =  10 m/s\\
Maximum flight speed in $z$-axis & $V_{z}$ =  5 m/s\\
Transmit power of GNs & $p_{k}$ =  30 dBm\\
Average channel power gain at 1 m & $\beta_{0}$ = 0 dB \\
Signal attenuation factor for NLoS & $\mu$ = --30 dB \\
Average path-loss exponent for LoS & $\alpha_{\textrm{L}}$ = 2 \\
Average path-loss exponent for NLoS & $\alpha_{\textrm{N}}$ = 2.7 \\
Noise power & $\sigma^{2}$ = --70 dBm \\
\hline
\end{tabular} 
\label{table1}
\end{center}
\end{table}

The parameters used for performance evaluations are shown in Table \ref{table1} \cite{Park23,Wu18,You20,Heo24,Heo24-2,Heo24-3,Kim24}. The GNs are distributed in a square area of size 100 m $\times$ 100 m, containing two buildings. The first building’s dimensions are $(\mathcal{W}_1,\mathcal{L}_1,\mathcal{H}_1)=(20,20,40)$ m, and the other's dimensions are $(\mathcal{W}_2,\mathcal{L}_2,\mathcal{H}_2)=(20,40,40)$ m. Additionally, the following parameters are used for the proposed algorithm implemented in Algorithm 1: $M$ = 300, $U$ = 10, $a^0$ = 0.01, $\varepsilon$ = 1.1, and $\epsilon$ = 0.0001.

For performance comparison, we consider the following four schemes.
\begin{enumerate}
     \item \emph{Proposed scheme}: The UAV strategy, including $\mathbf{S}$, $\mathbf{Q}$, and $\bar{\mathbf{C}}$, is optimized using Algorithm 1.
    
     \item \emph{LoS-based scheme}: The UAV optimizes $\mathbf{S}$ and $\mathbf{Q}$ based on a simplified LoS channel model, assuming all wireless channels are LoS \cite{Wu18}.

     \item \emph{Fixed altitude scheme}: The UAV altitude is fixed at 60 m higher than the height of buildings, and the UAV strategy, including $\mathbf{S}$, $\bar{\mathbf{C}}$, and horizontal trajectory, is optimized \cite{Yi24}.

     \item \emph{Fixed trajectory scheme}: The UAV operates in a hover-and-fly manner at an altitude of $H_{\textrm{min}}$, hovering above each GN sequentially and flying straight between them at maximum speed. Building avoidance and $\bar{\mathbf{C}}$ are predetermined along this fixed trajectory while optimizing $\mathbf{S}$.
\end{enumerate}

\begin{figure}[ht]
\centering
\includegraphics[width=0.9\linewidth]{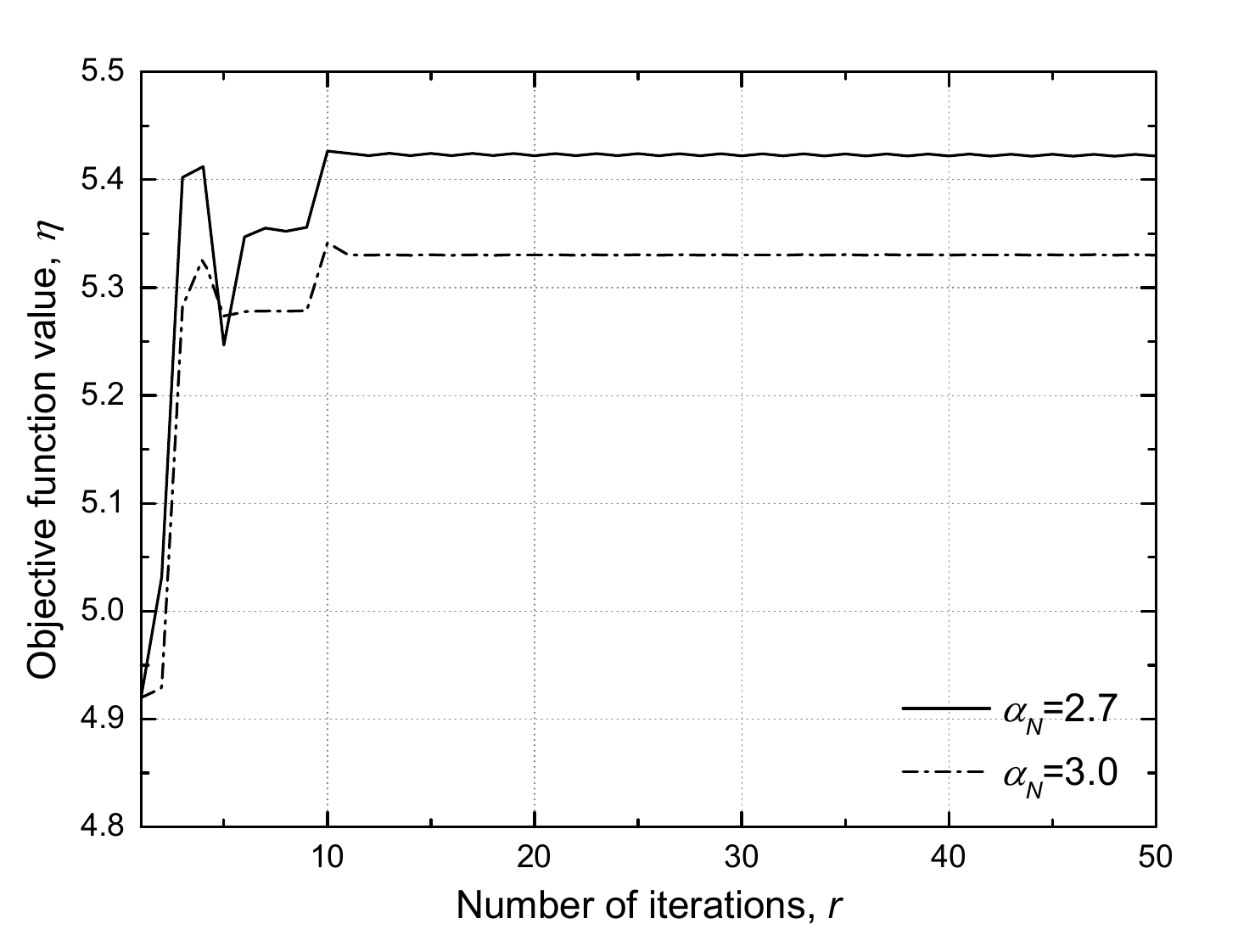} \caption{Convergence behavior of the proposed scheme.}
\label{R0}
\end{figure}

\begin{figure*}[ht!]
  \begin{center}
    \subfigure[3D trajectory of the proposed scheme.]{
      \includegraphics[width=0.315\linewidth]{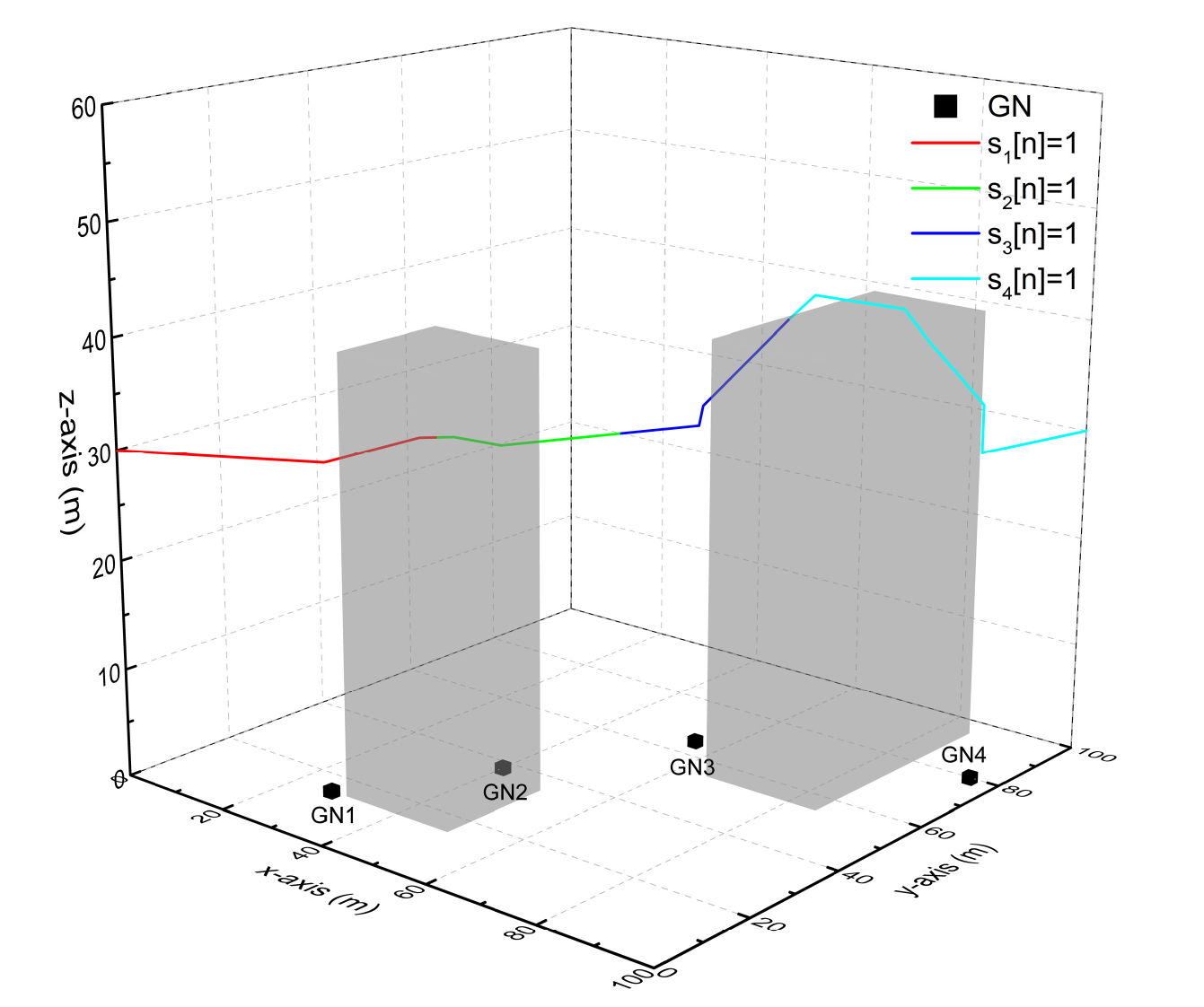}\label{R1-1}
    }
    \subfigure[Horizontal trajectory of the proposed scheme.]{
      \includegraphics[width=0.315\linewidth]{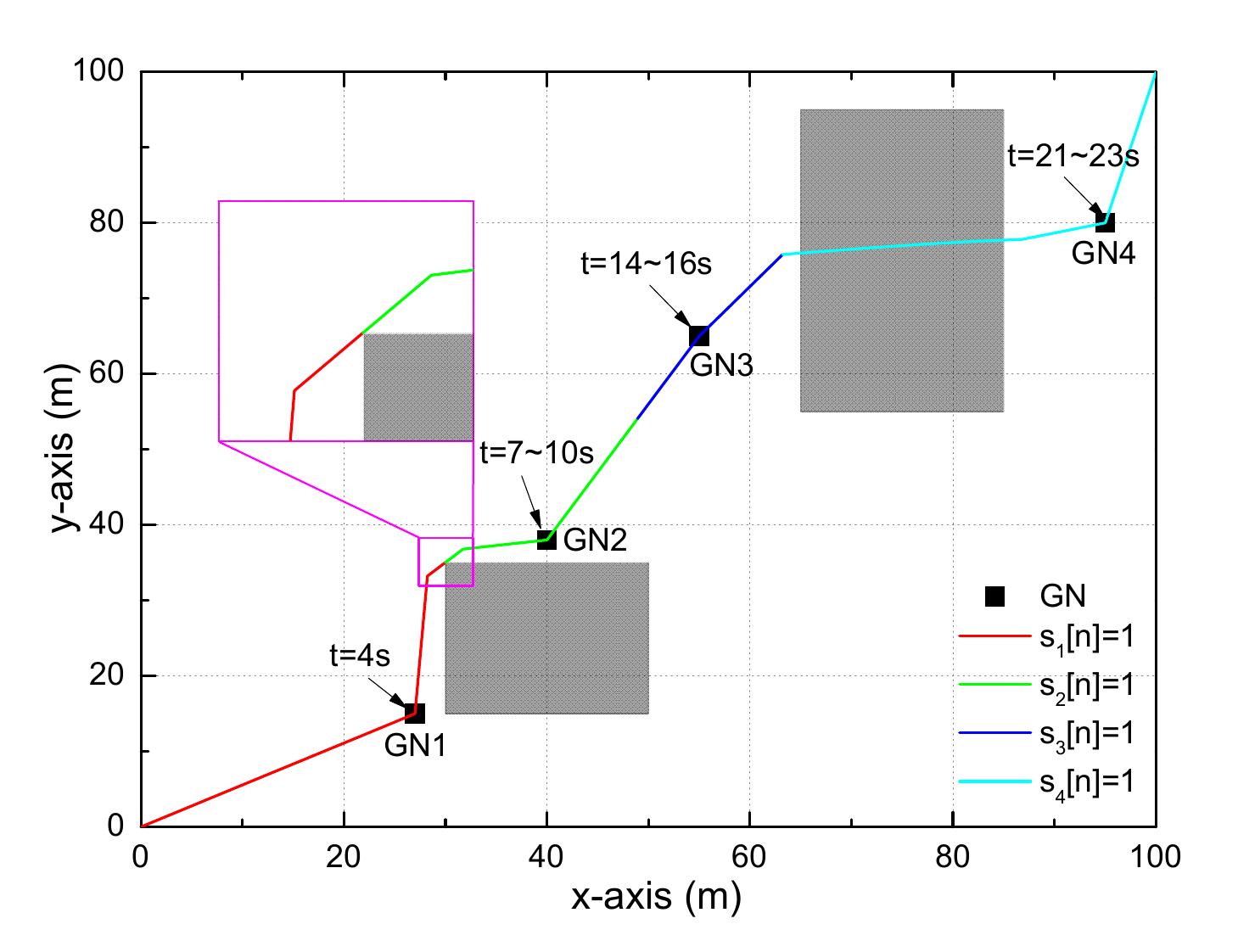}\label{R1-2}
    }
    \subfigure[Vertical trajectory of the proposed scheme.]{
      \includegraphics[width=0.315\linewidth]{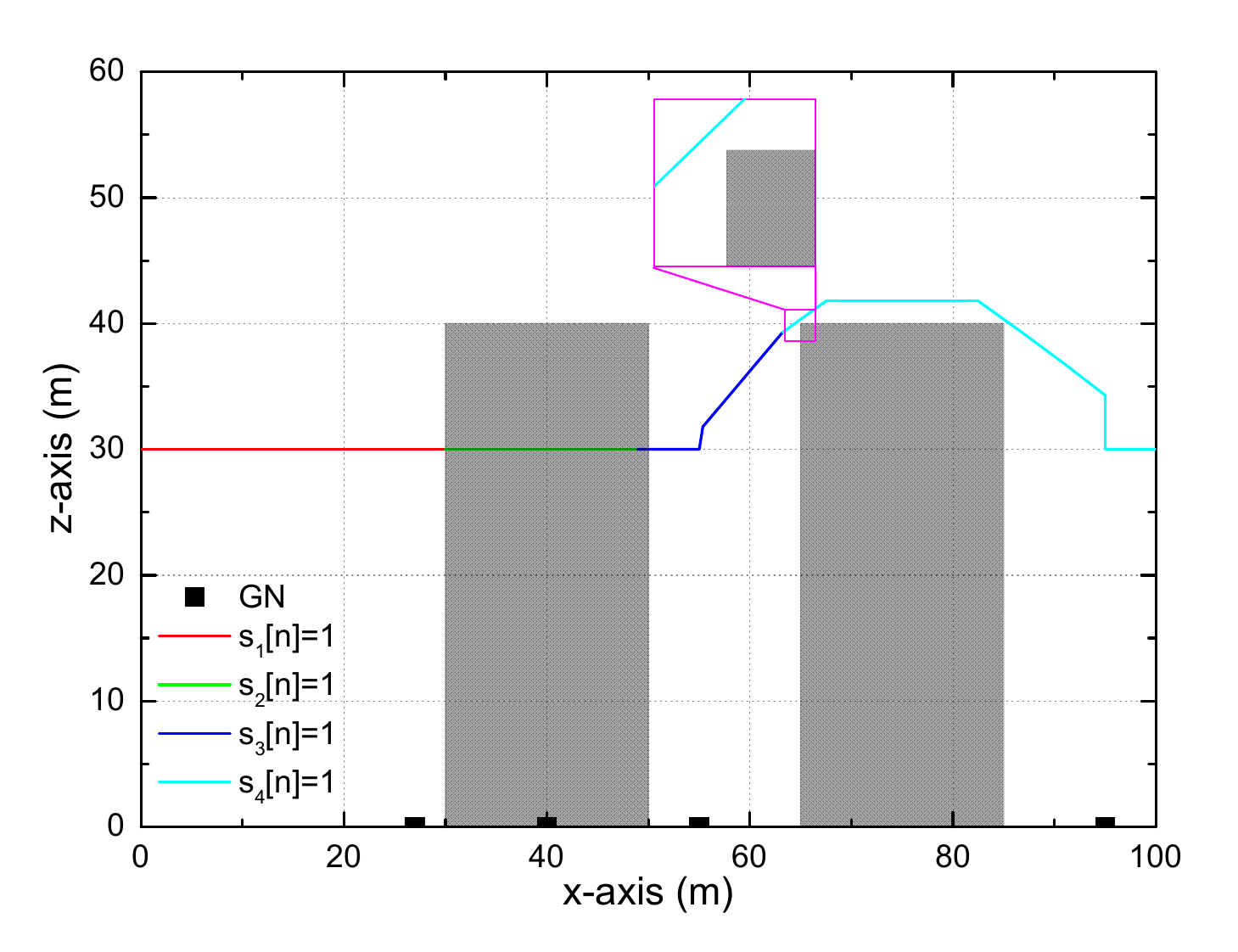}\label{R1-3}
    }
  \end{center} 
\caption{Scheduling and trajectory for the proposed scheme.}
\label{R1}
\end{figure*}

\begin{figure*}[ht!]
  \begin{center}
    \subfigure[Horizontal trajectory without the proposed building avoidance.]{
      \includegraphics[width=0.315\linewidth]{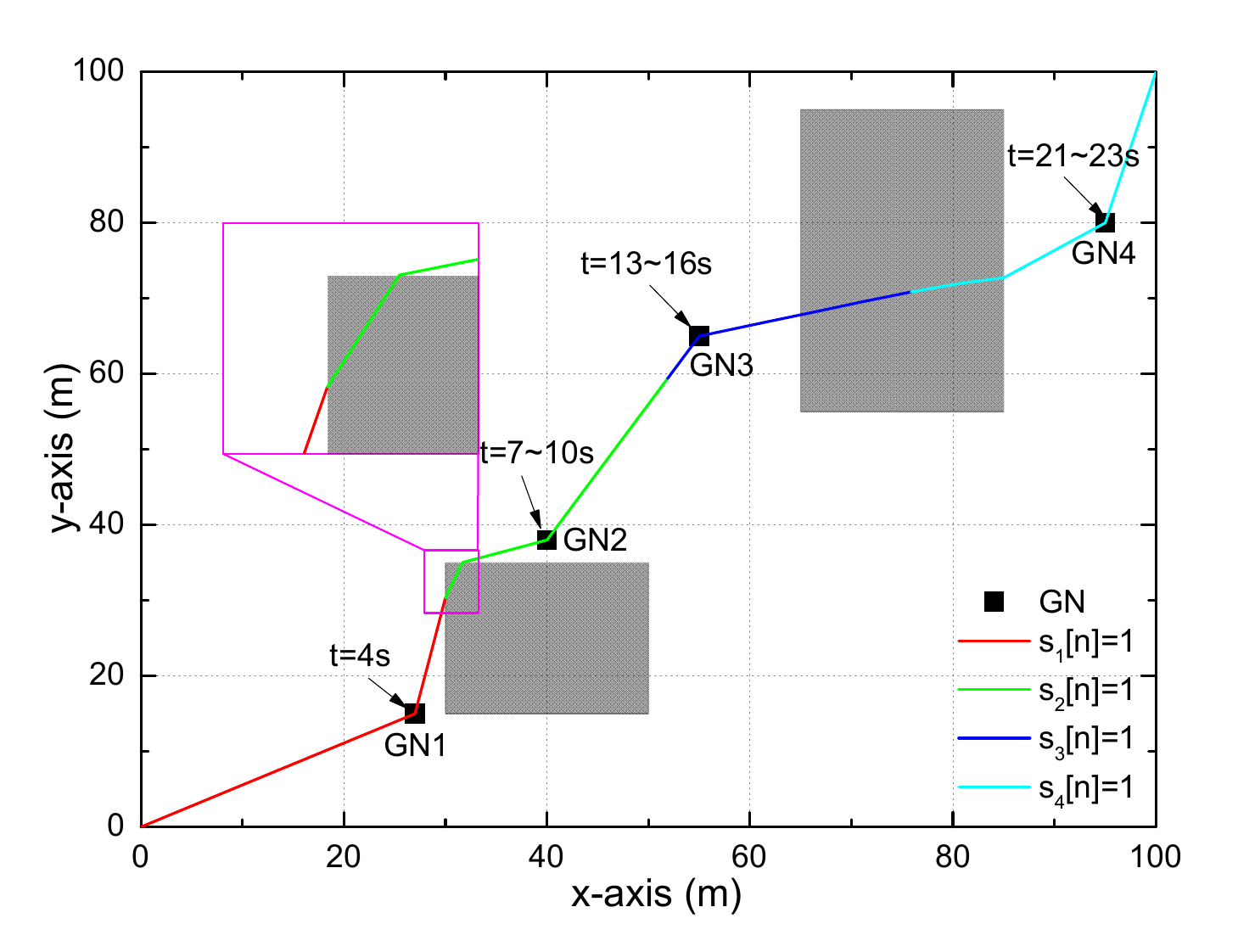}\label{R2-1}
    }
    \subfigure[Vertical trajectory without the proposed building avoidance.]{
      \includegraphics[width=0.315\linewidth]{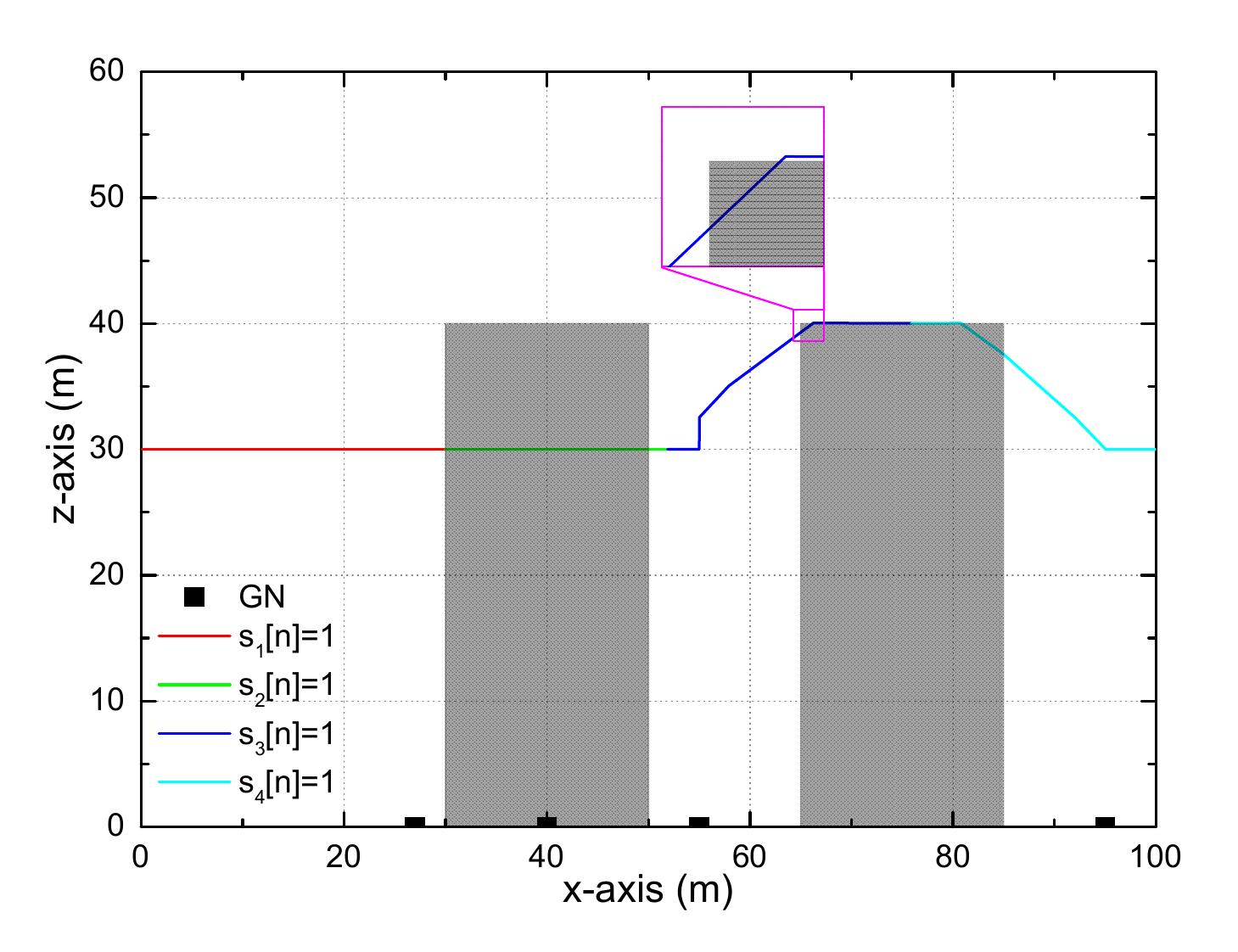}\label{R2-2}
    }
    \subfigure[LoS and NLoS regions for the proposed and LoS-based schemes.]{
      \includegraphics[width=0.315\linewidth]{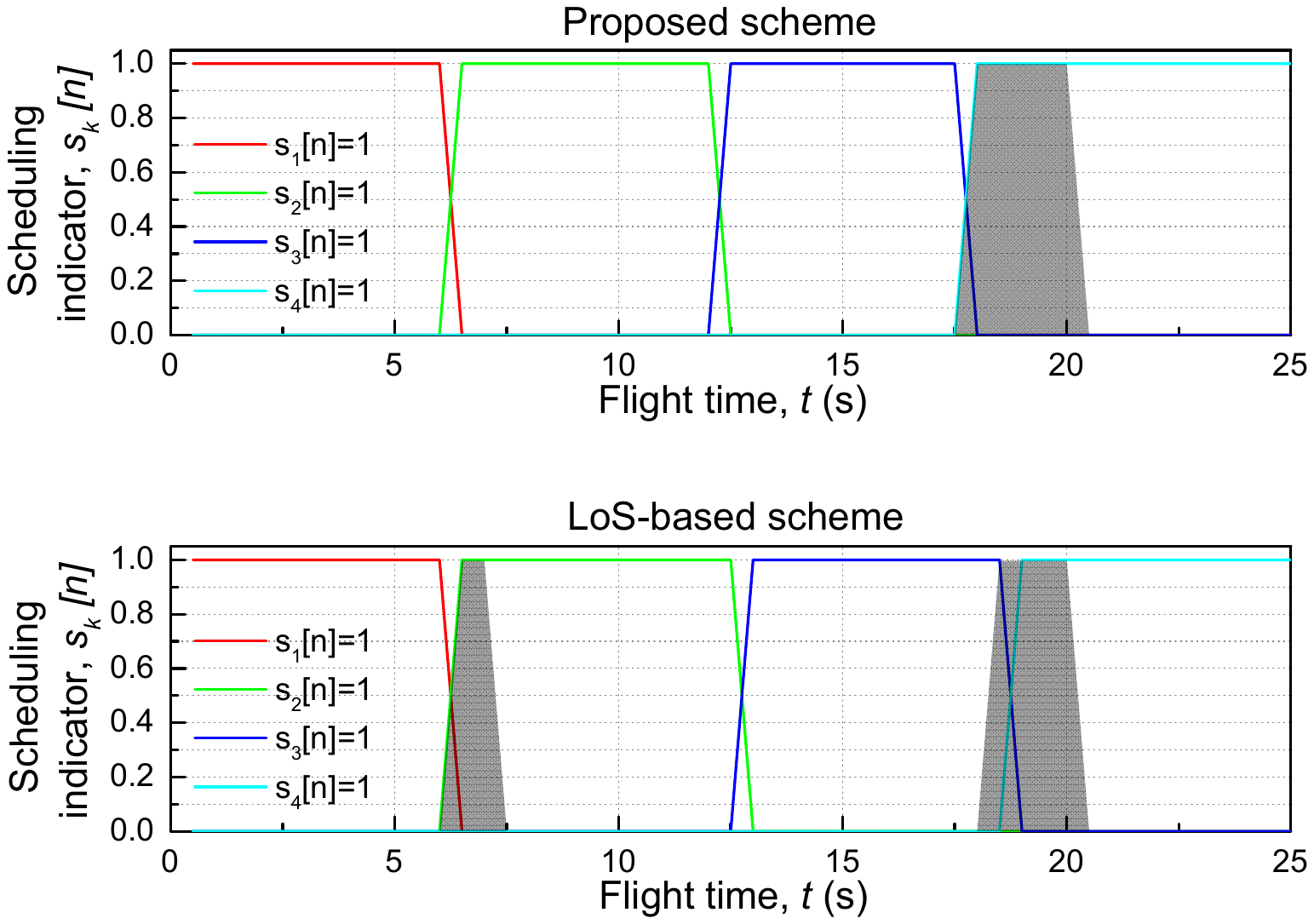}\label{R2-3}
    }
  \end{center} 
\caption{Impact of the proposed building avoidance and channel state determination.}
\label{R2}
\end{figure*}

Fig. \ref{R0} shows the convergence of the proposed scheme for varying $\alpha_{\textrm{N}}$. The objective function value ($\eta$) initially increases, decreases temporarily, and then increases again until convergence. This is because we use $\Phi_{a}^{\textrm{LB}}(\beta^{(i),u}_{k,l}[n])$ instead of the indicator function $\Phi(\beta^{(i),u}_{k,l}[n])$ while replacing \eqref{dsm61} with \eqref{indc2}. Initially, the small value of $a$ allows $\rho_{k,l}[n]$ to be optimized over a wider feasible region, leading to nonbinary values for $\rho_{k,l}[n]$ and an increase in $\eta$. As $a$ gradually increases, $\rho_{k,l}[n]$ converges to binary values, causing a temporary decrease in $\eta$. Subsequently, $\eta$ increases again and stabilizes as remaining optimization variables are optimized while satisfying the binary constraint of the LoS indicator. For $\alpha_{\textrm{N}}=2.7$, the signal attenuation due to the NLoS channel is less than $\alpha_{\textrm{N}}=3.0$, so $\eta$ has a higher value. Both cases converge in about $r=15$ iterations.

\begin{figure*}[ht!]
  \begin{center}
    \subfigure[$\bar{R}$ vs. $T$.]{
      \includegraphics[width=0.315\linewidth]{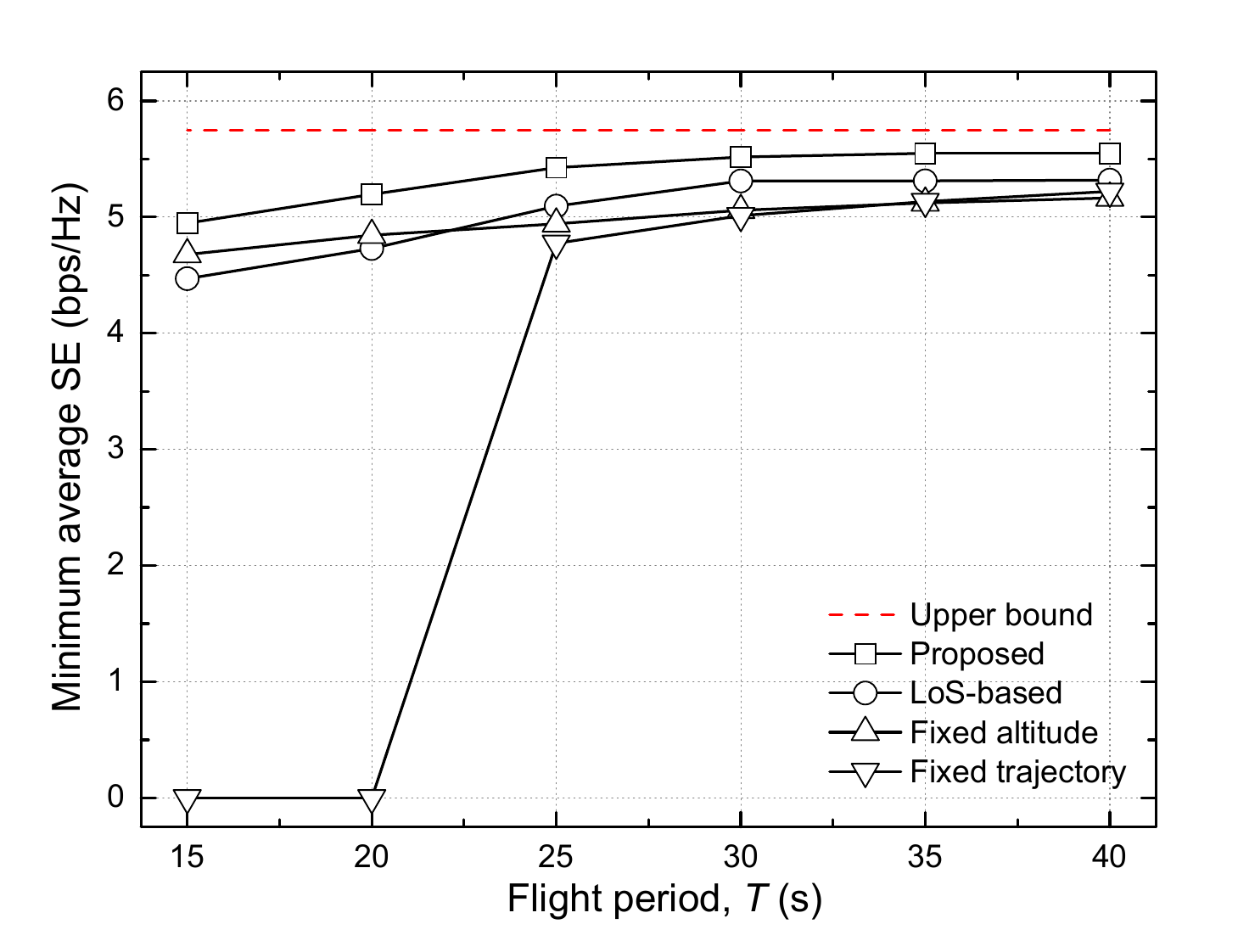}\label{R3-1}
    }
    \subfigure[$\bar{R}$ vs. $p_k$.]{
      \includegraphics[width=0.315\linewidth]{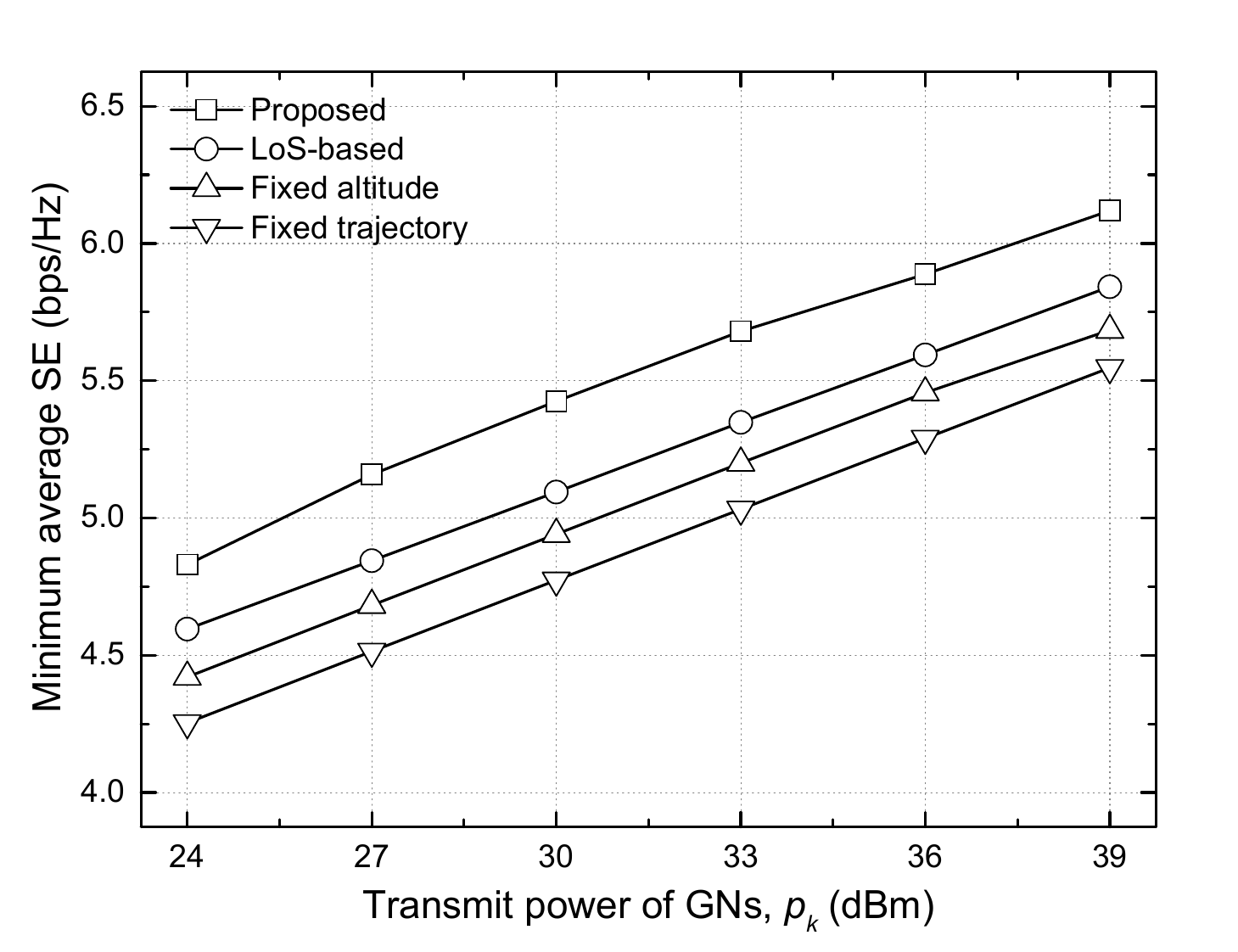}\label{R3-2}
    }
    \subfigure[$\bar{R}$ vs. $\alpha_{\textrm{N}}$.]{
      \includegraphics[width=0.315\linewidth]{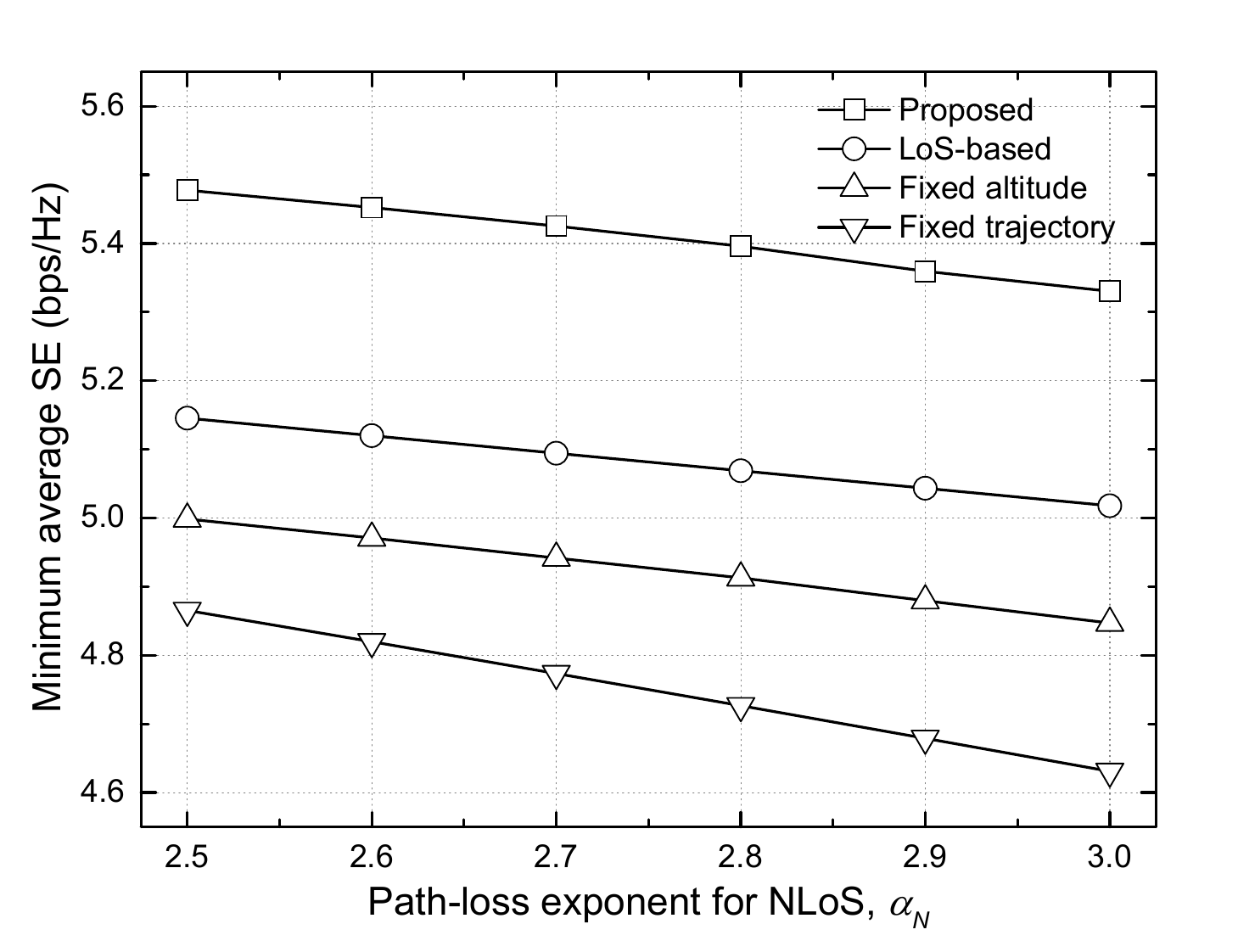}\label{R3-3}
    }
  \end{center} 
\caption{Performance comparison.}
\label{R3}
\end{figure*}

Fig. \ref{R1} shows the scheduling and trajectory for the proposed scheme. Fig. \ref{R1-1} shows the 3D trajectory of the UAV, with the scheduled GNs along the flight path in different colors. To aid interpretation, the horizontal and vertical trajectories are shown separately in Figs. \ref{R1-2} and \ref{R1-3}. As shown in Fig. \ref{R1-2}, the UAV flies directly over the scheduled GNs, staying for a certain period to effectively collect data. Moreover, the UAV makes a detour and flies next to the first building, which is narrower in length. In contrast, Fig. \ref{R1-3} shows that the UAV flies over the second building which is longer because avoiding it sideways would create an inefficient path. In both Figs. \ref{R1-2} and \ref{R1-3}, the proposed scheme considers an expanded building larger than its actual size for building avoidance, as described in \emph{Theorem 1}, and ensures a path that does not intersect the buildings, even during the UAV’s consecutive trajectory.

Fig. \ref{R2} illustrates the scheduling and trajectory for the LoS-based scheme with conventional building avoidance in \eqref{nffz}, validating the impact of the proposed building avoidance and channel state determination. Figs. \ref{R2-1} and \ref{R2-2} show that the UAV violates the buildings along its continuous trajectory, demonstrating the superiority of the proposed building avoidance method by comparing these results with Figs. \ref{R1-2} and \ref{R1-3}. Fig. \ref{R2-3} presents the LoS and NLoS regions for both the proposed and LoS-based schemes, with the grayed-out areas indicating where the NLoS channel is formed for the scheduled GNs. Combined with Fig. \ref{R1-3}, it is evident that the proposed scheme creates an NLoS channel with GN 4 only when the UAV flies over the second building to avoid it and the UAV allocates more time to GN 4 to compensate for the signal attenuation caused by the NLoS state. In contrast, in the LoS-based scheme, which does not consider signal blockage by buildings and assumes that the channel is always LoS, the NLoS channel is frequently formed with the scheduled GN due to incorrect scheduling policy and the UAV does not allocate additional time for GNs in NLoS conditions. As a result, in the proposed scheme, all GNs are served by the same SE $R_k=5.42$ bps/Hz, $\forall k$, while in the LoS-based scheme, the SE of GNs in NLoS channels is degraded, as shown by $(R_1,R_2,R_3,R_4)=(5.77, 5.36, 5.38, 5.05)$ bps/Hz, yielding a performance of $\min_{k\in\mathcal{K}} R_{k} \!=\! 5.05$ bps/Hz. This demonstrates the effectiveness of the proposed channel state determination.

Fig. \ref{R3} compares the performance in terms of the minimum average SE ($\bar{R}$) between the proposed and baseline schemes for (a) flight time ($T$), (b) transmit power of GNs ($p_k$), and (c) path-loss exponent for NLoS ($\alpha_{\textrm{N}}$). 

In Fig. \ref{R3-1}, the upper bound performance is achieved by removing the maximum velocity constraints, such as in \eqref{constM-3} and \eqref{constM-4}, and increasing $T$ until no further improvement is observed. As $T$ increases, the UAV's maneuverability has more degrees of freedom, allowing better optimization of its flight trajectory and radio resources. 
For example, the UAV can hover directly above the scheduled GNs for longer periods while maintaining the LoS channel, enabling data collection under more efficient channel conditions. As a result, increasing $T$ improves the minimum average SE for all schemes, and the proposed scheme approaches the upper bound performance. When $T$ is small, the UAV does not have enough time to avoid buildings, forcing the LoS-based scheme to fly at high altitudes to prevent all buildings, similar to the fixed altitude scheme. In this case, the LoS-based scheme optimizes the UAV strategy without considering the channel state, so even if the trajectories are similar, the GNs may be scheduled more frequently under NLoS conditions, resulting in performance degradation. 
However, as $T$ increases, the LoS-based scheme can avoid buildings at lower altitudes, improving channel conditions compared to the fixed altitude scheme. Thus, when $T \leq 20$ s, the LoS-based scheme performs worse than the fixed altitude scheme, but the performance reverses when $T \geq 25$ s. Similarly, for $T \leq 20$ s, the fixed trajectory scheme cannot optimize appropriately because the UAV cannot visit all GNs even at the maximum speed, rendering performance measurement impossible. As $T$ increases, the UAV is able to form a path that visits all GNs, resulting in a significant improvement in $\bar{R}$.

Fig. \ref{R3-2} shows that as $p_k$ increases, the strength of the signal transmitted by each GN increases, enabling the UAV to collect more data across all schemes. Moreover, Fig. \ref{R3-3} illustrates that as $\alpha_{\textrm{N}}$ increases, signal attenuation from the NLoS channel increases, leading to the degradation of $\bar{R}$ across all schemes. In these results, the best performance is observed in the order of LoS-based, fixed altitude, and fixed trajectory schemes among the baseline schemes. This suggests that optimizing horizontal trajectory, vertical trajectory, and channel state determination has the greatest impact on performance improvement in this order. Furthermore, the proposed scheme outperforms all baseline schemes by jointly optimizing scheduling and 3D trajectory and accurately determining the LoS and NLoS state for wireless channels.

\section{Conclusion}

This study leveraged UAV mobility to address signal blockage and building avoidance in urban environments. 
A mathematical model, supported by rigorous mathematical proofs, was developed to determine wireless signal blockage caused by cuboid-shaped buildings, ensuring that UAV trajectories do not encroach on these buildings. 
Using this model, we formulated a joint optimization problem of scheduling and 3D trajectory while accurately determining the LoS/NLoS channel state for UAV-assisted data harvesting to maximize the minimum uplink throughput among GNs. We also employed various optimization techniques to solve the nonconvex MINLP problem, such as QT, SCA, BCD, the separating hyperplane method, and an approximated indicator function. Through comprehensive simulations, we demonstrated that the UAV can effectively adjust its trajectory and scheduling policy to maintain LoS channels with scheduled GNs, thereby enhancing uplink throughput while avoiding cuboid-shaped buildings. We expect that this study will contribute to improving UAV communication performance in urban areas by dynamically leveraging LoS channels and ensuring building avoidance.

\section*{Appendix}

then the segment length satisfies is less than

Consider two cuboids with the same center coordinates at $(0,0,0)$: one cuboid $B$ has half-width $\frac{\mathcal{W}_B}{2}$, half-length $\frac{\mathcal{L}_B}{2}$, and half-height $\frac{\mathcal{H}_B}{2}$, with representing $\mathcal{B}$ be the set of the interior points of $B$. The other cuboid $B_{\textrm{exp}}$ has expanded dimensions with half-width $\frac{\mathcal{W}_B}{2}+\frac{d_{\textrm{max}}}{2\sqrt{2}}$, half-length $\frac{\mathcal{L}_B}{2}+\frac{d_{\textrm{max}}}{2\sqrt{2}}$, and half-height $\frac{\mathcal{H}_B}{2}+\frac{d_{\textrm{max}}}{2\sqrt{2}}$ where $d_{\textrm{max}} < \min(\mathcal{W}_B,\mathcal{L}_B,\mathcal{H}_B)$, and let $\mathcal{B}_{\textrm{exp}}$ be the set of the interior points of $B_{\textrm{exp}}$, therefore, $\mathcal{B} \subset \mathcal{B}_{\textrm{exp}}$ holds. We assume that if a line segment connects any two points that are not contained in $\mathcal{B}_{\textrm{exp}}$, i.e., $\mathbf{q}_1$, $\mathbf{q}_2$ $\notin \mathcal{B}_{\textrm{exp}}$, then the segment length satisfies $\|\overline{\mathbf{q}_1\mathbf{q}_2}\| \leq d_{\textrm{max}}$. 

There are two cases: i) when the line segment $\overline{\mathbf{q}_1\mathbf{q}_2}$ does not include any point $\mathbf{p}$ contained in $\mathcal{B}_{\textrm{exp}}$, i.e.,  $\overline{\mathbf{q}_1\mathbf{q}_2}\cap \mathcal{B}_{\textrm{exp}} = \emptyset$, and ii) when it does, i.e.,  $\overline{\mathbf{q}_1\mathbf{q}_2}\cap \mathcal{B}_{\textrm{exp}} \neq \emptyset$.  

In the first case, it is always guaranteed that $\overline{\mathbf{q}_1\mathbf{q}_2}$ does not include any point in $\mathcal{B}$ because $\mathcal{B} \subset \mathcal{B}_{\textrm{exp}}$. 

In the second case, there are two points $\mathbf{q}_1'$ and $\mathbf{q}_2'$ that intersect the line segment $\overline{\mathbf{q}_1\mathbf{q}_2}$ with the faces forming the cuboid $B_{\textrm{exp}}$. Since $\overline{\mathbf{q}_1\mathbf{q}_2}$ includes $\mathbf{p}$, the interior point of $B_{\textrm{exp}}$, it is obvious that $\mathbf{q}_1' \neq \mathbf{q}_2'$ and $\|\overline{\mathbf{q}_1'\mathbf{q}_2'}\| \leq d_{\textrm{max}}$ are satisfied. Therefore, $\mathbf{q}_1'$ and $\mathbf{q}_2'$ each belong to one of the six faces of the cuboid $B_{\textrm{exp}}$. Subsequently, there are three possible subcases: ii-1) when $\mathbf{q}_1'$ and $\mathbf{q}_2'$ belong to the parallel faces, ii-2) when they belong to the same face, and iii-3) when they belong to the adjacent faces. 

Because of the condition $\|\overline{\mathbf{q}_1\mathbf{q}_2}\| \leq d_{\textrm{max}}$, subcase ii-1) cannot occur. For subcase ii-2), every point of $\overline{\mathbf{q}_1'\mathbf{q}_2'}$ is not contained in $\mathcal{B}$ because any point on the faces of $B_{exp}$ cannot be included in $\mathcal{B}$. Finally, we must consider subcase ii-3). We note that each face of the cuboid has the same component for one axis, so if a point belongs to face $\mathcal{F}$, we know the axis component perpendicular to that face. For example, the $z$-axis component of a point belonging to face $\mathcal{F}_z$ that lies in the direction of the positive z-axis perpendicular to the z-axis is $\frac{\mathcal{H}_B}{2} + \frac{d_{\textrm{max}}}{2\sqrt{2}}$. Here, we consider the case where $\mathbf{q}_1'$ and $\mathbf{q}_2'$ belong to $\mathcal{F}_z$ and $\mathcal{F}_y$ of $B_{\textrm{exp}}$, respectively, and then, their respective coordinates are $\mathbf{q}'_1 \!=\! \left(\!x_1, y_1, \frac{\mathcal{H}_B}{2}\!+\!\frac{d_{\textrm{max}}}{2\sqrt{2}}\!\right)$ and $\mathbf{q}'_2 \!=\! \left(\!x_2, \frac{\mathcal{L}_B}{2}\!+\!\frac{d_{\textrm{max}}}{2\sqrt{2}}, z_2\!\right)$. Notably, the same proof can be applied even if any other two adjacent faces are selected.

We will prove by contradiction that the line segment $\overline{\mathbf{q}_1'\mathbf{q}_2'}$ does not violate the interior points of the cuboid $B$. Assume $\overline{\mathbf{q}_1'\mathbf{q}_2'}$ invades $\mathcal{B}$, i.e. $\overline{\mathbf{q}_1'\mathbf{q}_2'} \cap \mathcal{B} \neq \emptyset$, and let $\mathbf{v}=(v_x,v_y,v_z)$ be the point included in $\overline{\mathbf{q}_1'\mathbf{q}_2'} \cap \mathcal{B}$. In other words, $\mathbf{v}$ can be considered as an internal division point of $\overline{\mathbf{q}_1'\mathbf{q}_2'}$ as well as the element of $\mathcal{B}$. For $\mathbf{v}$ to be the element of $\mathcal{B}$, it must satisfy the following conditions: 
\begin{subequations}
\begin{align}
    -\frac{\mathcal{W}_B}{2} &< v_x < \frac{\mathcal{W}_B}{2},\label{ix_ineq}\\
     -\frac{\mathcal{L}_B}{2} &< v_y < \frac{\mathcal{L}_B}{2},\label{iy_ineq}\\
     -\frac{\mathcal{H}_B}{2} &< v_z < \frac{\mathcal{H}_B}{2}.\label{iz_ineq}
\end{align}
\end{subequations}
Moreover, since $\mathbf{v}$ is the internal division point of $\overline{\mathbf{q}_1'\mathbf{q}_2'}$, its components can be expressed as 
\begin{subequations}
\begin{align}
    v_x &= x_1 + (x_2-x_1)t, \\
    v_y &= y_1 + \bigg(\frac{\mathcal{L}_B}{2}+\frac{d_{\textrm{max}}}{2\sqrt{2}} - y_1\bigg)t \nonumber\\
    &=\frac{\mathcal{L}_B}{2}+\frac{d_{\textrm{max}}}{2\sqrt{2}} + \bigg(y_1 -\frac{\mathcal{L}_B}{2}-\frac{d_{\textrm{max}}}{2\sqrt{2}}\bigg)(1-t), \label{vy}\\
    v_z &= \frac{\mathcal{H}_B}{2}+\frac{d_{\textrm{max}}}{2\sqrt{2}} + \bigg(z_2 -\frac{\mathcal{H}_B}{2}-\frac{d_{\textrm{max}}}{2\sqrt{2}}\bigg)t, \label{vz}
\end{align}
\end{subequations}
where $0 < t < 1$.

Substituting \eqref{vy} and \eqref{vz} into \eqref{iy_ineq} and \eqref{iz_ineq}, we obtain the following inequalities:
\begin{align}
    \frac{-\mathcal{L}_B-\frac{d_{\textrm{max}}}{2\sqrt{2}}}{1-t}<  y_1 - \frac{\mathcal{L}_B}{2} - \frac{d_{\textrm{max}}}{2\sqrt{2}}  < \frac{-\frac{d_{\textrm{max}}}{2\sqrt{2}}}{1-t},\label{iy_ineq_t}\\
    \frac{-\mathcal{H}_B-\frac{d_{\textrm{max}}}{2\sqrt{2}}}{t}< z_2 - \frac{\mathcal{H}_B}{2} - \frac{d_{\textrm{max}}}{2\sqrt{2}}  < \frac{- \frac{d_{\textrm{max}}}{2\sqrt{2}}}{t}. \label{iz_ineq_t}
\end{align}
Since each LHS of \eqref{iy_ineq_t} and \eqref{iz_ineq_t} are always negative values, the following inequalities can be constructed.
\begin{align}
    \bigg( y_1 - \frac{\mathcal{L}_B}{2} - \frac{d_{\textrm{max}}}{2\sqrt{2}} \bigg)^2 &> \frac{d_{\textrm{max}}^2}{8(1-t)^2},\label{y1_square} \\
    \bigg(z_2 - \frac{\mathcal{H}_B}{2} - \frac{d_{\textrm{max}}}{2\sqrt{2}} \bigg)^2 &> \frac{d_{\textrm{max}}^2}{8t^2}.\label{z1_square}
\end{align}
By adding \eqref{y1_square} and \eqref{z1_square}, we can get the following inequalities: 
\begin{align}
     &\bigg( y_1 - \frac{\mathcal{L}_B}{2} - \frac{d_{\textrm{max}}}{2\sqrt{2}} \bigg)^{2} + \bigg(z_2 - \frac{\mathcal{H}_B}{2} - \frac{d_{\textrm{max}}}{2\sqrt{2}} \bigg)^{2} \nonumber \\
    &~~~~~~~~~~~~~~~~~~~~~~~> \frac{d_{\textrm{max}}^2}{8}\bigg(\frac{1}{t^2} + \frac{1}{(1\!-\!t)^2} \bigg) \overset{(a)}{\geq} d_{\textrm{max}}^2. \label{aa}
\end{align}
Here, inequality ($a$) in \eqref{aa} holds because $f(t)=\frac{1}{t^2} + \frac{1}{(1\!-\!t)^2}$ is convex w.r.t. $t$ for $0 < t < 1$ and the minimum value of $f(t)$ is $8$ at $t=0.5$. As a result, the length of $\overline{\mathbf{q}_1'\mathbf{q}_2'}$ becomes larger than $d_{\textrm{max}}$, as follows:
\begin{align}
    \|\overline{\mathbf{q}_1'\mathbf{q}_2'}\|^2 &= (x_2 - x_1)^2 + \bigg( y_1 - \frac{\mathcal{L}_B}{2} - \frac{d_{\textrm{max}}}{2\sqrt{2}} \bigg)^2 \nonumber \\
    & ~~~~+ \bigg(z_2 - \frac{\mathcal{H}_B}{2} - \frac{d_{\textrm{max}}}{2\sqrt{2}} \bigg)^2 > d_{\textrm{max}}^2. 
\end{align}
This means that if $\overline{\mathbf{q}_1'\mathbf{q}_2'}$ violates the interior points of the cuboid $B$, then the original assumption that $\|\overline{\mathbf{q}_1\mathbf{q}_2}\| \leq d_{\textrm{max}}$ does not hold because of $\|\overline{\mathbf{q}_1'\mathbf{q}_2'}\| \leq \|\overline{\mathbf{q}_1\mathbf{q}_2}\|$, which is a contradiction. 

Therefore, we conclude that if points $\mathbf{q}_1$ and $\mathbf{q}_2$ are not contained in $\mathcal{B}_{\textrm{exp}}$, the line segment $\overline{\mathbf{q}_1\mathbf{q}_2}$ never violates the cuboid $B$.


\end{document}